\documentclass[11pt,a4paper]{article}
\pdfoutput=1

\usepackage{jcappub}

\usepackage[T1]{fontenc}
\usepackage{lmodern}        
\usepackage[a4paper,top=2.8cm,bottom=2.6cm,left=1.8cm,right=1.8cm]{geometry}
\usepackage{graphicx}
\usepackage{amsmath,amssymb}
\usepackage{float}
\usepackage{xcolor}

\usepackage{url}       
\usepackage{doi}     
\usepackage[numbers,sort&compress]{natbib}
\usepackage{hyperref}  

\hypersetup{
  colorlinks = true,
  linkcolor = red,
  urlcolor  = magenta,
  citecolor = blue,
  anchorcolor = blue
}

\title{Non-linear structure formation with elastic interactions in the dark sector}

\author[1,4]{Jose Beltr\'an Jim\'enez,}
\author[2,4]{David Figueruelo,}
\author[4]{David F. Mota,}
\author[4]{Hans A. Winther}

\affiliation[1]{Departamento de F\'isica Fundamental and IUFFyM, Universidad de Salamanca, 37008 Salamanca, Spain}
\affiliation[2]{Department of Theoretical Physics, University of the Basque Country UPV/EHU, 48080 Bilbao, Spain}
\affiliation[4]{Institute of Theoretical Astrophysics, University of Oslo, N-0315 Oslo, Norway}

\emailAdd{jose.beltran@usal.es}
\emailAdd{david.figueruelo@ehu.eus}
\emailAdd{d.f.mota@astro.uio.no}
\emailAdd{h.a.winther@astro.uio.no}

\abstract{
Cosmological models where dark matter interacts with dark energy via a pure momentum transfer and with no energy exchange (i.e. elastic) provide compelling scenarios for addressing the apparent lack of structures at low redshift. In particular, it has been shown that measurements of $S_8$ may show a statistically significant preference for the presence of elastic interactions. In this work we implement a specific realisation of these scenarios into an $N$-body code to explore the non-linear regime. We include two populations of particles to describe the interacting dark matter and the non-interacting baryons respectively. On linear scales we recover the suppression of structures obtained from Boltzmann codes, while non-linear scales exhibit an enhancement of the matter power. We find that fewer massive halos are formed at low redshift as a consequence of the elastic interaction and that dark matter halos are more compact than in the standard model. Furthermore, the ratio of dark matter and baryons density profiles is not constant. Finally, we corroborate that baryons efficiently cluster around dark matter halos so they provide good tracers of the dark matter velocity field despite the presence of the interaction. This shows that the interaction is not sufficiently strong as to disrupt virialised structures.
}

\keywords{Dark matter, dark energy, interactions, non-linear structure formation}

\begin{document}
\maketitle

\section{Introduction}
The advances in observational cosmology over recent decades have firmly established the $\Lambda$CDM model as the standard paradigm. From Cosmic Microwave Background (CMB) data by experiments like Planck \cite{Aghanim:2018eyx} to Large Scale Structure (LSS) surveys like SDSS  \cite{eBOSS:2020yzd} or DES \cite{DES:2021wwk}, the establishment of a Universe dominated by the Dark Sector has consolidated the $\Lambda$CDM  paradigm. Despite the successes of $\Lambda$CDM in describing a wide range of cosmological observations, the true nature of the dark sector remains elusive, and its fundamental constituents continue to challenge our theoretical understanding. Moreover, persistent tensions between different experiments suggest that new physics may be required to fully account for the formation of cosmic structures or the expansion of the Universe. In particular, the $H_0$ crisis and the $\sigma_8$, or $S_8$, tension (see e.g. \cite{Perivolaropoulos:2021jda,Abdalla:2022yfr} for a review) stand out as the most pressing discrepancies in the current cosmic arena, while recently the new results of DESI experiment has raised the question on whether dark energy has a dynamical nature~\cite{DESI:2025zgx}. Alternatively, the previous discrepancies can be explained by not yet accounted systematics in the observations of the cosmic standard candles or rulers~\cite{CosmoVerseNetwork:2025alb}. Whether these tensions arise from systematics or hint at new physics stands as a key question in modern cosmology.

Considering the latter possibility, and among all the plethora of possibilities, elastic interactions within the Dark Sector consisting of a momentum exchange between dark energy and dark matter have emerged as a promising avenue to alleviate the observed discrepancies in the $\sigma_8$ tension~\cite{Simpson:2010vh,Amendola:2020ldb,Pourtsidou:2016ico,Kumar:2017bpv,Chamings:2019kcl,Vagnozzi:2019kvw,Linton:2021cgd,Ferlito:2022mok}. In these scenarios, the cosmological constant $\Lambda$ is replaced by a dynamical dark energy which is coupled to a matter component of the Universe. The coupling is in the form of a pure momentum exchange between dark energy and the matter component, preventing the matter from continuing to cluster as it would in the standard scenario. Essentially, dark energy exerts a drag on dark matter, suppressing its ability to form structures and, as a consequence, the overall growth of cosmic structures is reduced. While multiple models exist that incorporate this kind of momentum transfer~\cite{Simpson:2010vh,Pourtsidou:2013nha,Skordis:2015yra,Koivisto:2015qua,Kase:2019mox,Chamings:2019kcl,Amendola:2020ldb,Pourtsidou:2016ico,Kumar:2017bpv,Nakamura:2019phn,Vagnozzi:2019kvw,DeFelice:2020icf,Linton:2021cgd,ManciniSpurio:2021jvx,Ferlito:2022mok,Cardona:2023gzq,Pookkillath:2024ycd,Aoki:2025bmj}, here we will examine the so-called covariantised dark Thomson-like scattering as a proxy for understanding how the non-linear clustering is modified by the presence of an elastic interaction. This particular scenario features a straightforward and phenomenological formulation, with only one new parameter governing both the strength of the interaction and the time at which it becomes relevant. For this reason, it represents a convenient proxy for studying the broad class of momentum transfer models. It was first introduced in \cite{Asghari:2019qld} and later extensively analysed in \cite{Figueruelo:2021elm,BeltranJimenez:2021wbq,Cardona:2022mdq,BeltranJimenez:2022irm,BeltranJimenez:2024lml} (see \cite{BeltranJimenez:2024lml} for a review). Among the most remarkable features of this scenario is its ability to alleviate the $\sigma_8$ tension with just one extra parameter \cite{Asghari:2019qld,Figueruelo:2021elm}, a result that was later reproduced by other authors in \cite{Poulin:2022sgp}, and its signature in the dipole of the matter power spectrum \cite{BeltranJimenez:2022irm} that could be detected with future SKA-like surveys and, thus, serve as a smoking gun for these models. The philosophy behind the covariantised dark Thomson-like scattering involves an interaction between dark energy and a matter fluid analogous to the Thomson scattering that occurred before recombination. During that era, dark matter had already decoupled, while the remaining components formed a plasma where photons and electrons were coupled through Thomson scattering. As a result, baryons were unable to collapse, since the radiation pressure from photons counteracted the gravitational collapse of baryons, preventing the formation of gravitationally bound structures. In this framework, the elastic interaction between dark energy and a matter fluid operates similarly. Here, the pressure from dark energy can counterbalance the gravitational collapse of the coupled matter component while the interaction is efficient, that is while the fluid with pressure is relevant. Consequently, since the fluid with pressure, dark energy, only appears in the cosmic arena at late times, the covariantised dark Thomson-like scattering  belongs to the late Universe\footnote{For Thomson scattering the time-scale when the interaction was efficient is the early Universe since then was when the pressure-bearing component, photons, was non-negligible.}. In this work, we will focus on the interaction between dark energy and dark matter, while baryons remain uncoupled. This allows us to specifically study how the exchange of momentum between dark energy and dark matter influences cosmic structure formation without introducing additional complexities related to baryonic interactions. The linear analysis of the same interaction, but with baryons, was performed in \cite{BeltranJimenez:2020iyx} and motivated by the elastic dark energy-baryons interaction considered in \cite{Vagnozzi:2019kvw}. \\

This paper is organised as follows: in section~\ref{sec:model} we present the theory of the covariantised dark Thomson-like scattering, in section~\ref{sec:non-linear} we present the non-linear prescription and details of the N-body analysis performed, in section~\ref{sec:results} we show the main results obtained for the matter power spectrum, the halo formation process and halo profiles, together with the effects of the interaction on the cosmic web and finally, in section~\ref{sec:conclusion} we summarise the final conclusions of this work. 

\section{The elastic interacting model}
\label{sec:model}

The covariantised dark Thomson-like scattering model is based on a modification of the energy-momentum conservation equations of the dark sector expressed as:
\begin{eqnarray}
    \nabla_\mu T^{\mu \nu}_{\rm dm} &=& Q^\nu\;, \label{eq:non-conserv_a}\\
    \nabla_\mu T^{\mu \nu}_{\rm de} &=& -Q^\nu\;,
    \label{eq:non-conserv_b}
\end{eqnarray}
where $Q^\nu$ encodes the interaction and it is defined as
\begin{eqnarray}
     Q^{\nu} &\equiv&  \bar{\alpha}(u^\nu_{\rm de} - u^\nu_{\rm dm})\;,
     \label{eq:defQ}
\end{eqnarray}
with $u_i^\nu$ the 4-velocity of the $i$-th component. The amount of momentum transfer and therefore the strength of the interaction is controlled by the coupling parameter $\bar{\alpha}$, which will also determine the time when the  interaction becomes relevant. In order to work with dimensionless couplings, we conveniently normalise $\bar{\alpha}$ as
\begin{eqnarray}
    \alpha = \frac{8 \pi G}{3 H_0^3} \bar{\alpha}\;.
\end{eqnarray}
Since the covariantised dark Thomson-like interaction is proportional to the relative velocity of the fluids, the background evolution remains unaffected. In the comoving rest-frame of an isotropic and homogeneous Universe, the 4-velocities of all the fluid components are identical so that we have
\begin{eqnarray}
Q^\nu\Big|_{\textnormal{background}} \propto \left(u^\nu_{\rm de} - u^\nu_{\rm dm}\right)\Big|_{{\textnormal{background}}} = 0\;. 
\end{eqnarray}
Thus, the background cosmology, Friedmann equations, continuity equations, and the Hubble function, remain as in the standard model and all the effects only appear in the perturbations. The only caveat is that a pure $\Lambda$CDM model cannot be used for the background because the interaction depends on the peculiar velocities of dark energy and a cosmological constant, being a constant, does not have perturbations. Thus, we need to consider some dynamical dark energy model that we will take for simplicity as $w$CDM here, i.e., dark energy in the form of a perfect fluid with constant equation of state parameter $w$.

On small scales and at late times when there are non-negligible peculiar velocities between dark matter and dark energy, the interaction will commence to affect the evolution of the perturbations. To see how this comes about, let us consider scalar perturbations in the Newtonian gauge where the metric is described by the following line element: 
\begin{equation}
{\rm d} s^2=a^2(\tau)\left[-\big(1+2\Phi\big){\rm d} \tau^2 +\big(1-2\Psi\big){\rm d}\vec{x}^2\right]\,.
\end{equation}
Since we will not consider anisotropic stresses, the gravitational potentials will satisfy $\Psi=\Phi$. In this gauge, the velocity perturbation at first order takes the form 
\begin{eqnarray}
u^\mu = \frac{1}{a}\big(1-\Phi, \vec{v}\big)\;,
\end{eqnarray}
where $\vec{v} = \tfrac{{\rm d} \vec{x}}{{\rm d} \tau}$ describes the peculiar velocities.  
We can now use equations~\eqref{eq:non-conserv_a} and~\eqref{eq:non-conserv_b} to obtain the equations governing the linear evolution of the density contrast $\delta\equiv \frac{\delta\rho}{\rho}$ and the velocity perturbations in Fourier space defined as $\theta\equiv i \Vec{k} \cdot\vec{v}$. Their explicit expressions can be written as
\begin{eqnarray}
\label{eq:deltaDM}
\delta_{\rm dm}' &=& 
-\theta_{\rm dm}+3\Phi'\,,\\
\label{eq:deltaDE}
\delta_{\rm de}'&=&-3 \mathcal{H}( c_{\rm de}^2-w) \delta_{\rm de} +3(1+w)\Phi'  -(1+w)\left(1+9 \mathcal{H}^2
\frac{c_{\rm de}^2-w}{k^2}\right)
\theta_{\rm de}\;,  \\
\label{eq:thetaDM}
\theta_{\rm dm}'&=&-\mathcal{H} \theta_{\rm dm} + k^2 \Phi + \Gamma(\theta_{\rm de}-\theta_{\rm dm})\;, \\
\label{eq:thetaDE}
\theta_{\rm de}' &=&(3c_{\rm de}^2-1) \mathcal{H}\theta_{\rm de}+k^2\Phi +\frac{k^2 c_{\rm de}^2}{1+w}\delta_{\rm de}-\Gamma R(\theta_{\rm de}-\theta_{\rm dm})\,,
\end{eqnarray}
where $w$ is the dark energy equation of state parameter, $c_{\rm de}^2$ its sound speed squared and $\mathcal{H}=a'/a$ is the conformal Hubble function. In the above equations, we have defined the quantities $\Gamma$ and $R$ as follows:
\begin{eqnarray}
\Gamma&\equiv& \alpha \frac{a^4}{\Omega_{\rm dm}} \;, \\ \label{eq:Scoupling}
R &\equiv&
\frac{\rho_{\rm dm}}{(1+w)\rho_{\rm de}}\,. \label{eq:Rcoupling}
\end{eqnarray}
These quantities represent the effective interaction rate and the relative density of dark matter to dark energy and they govern the importance of the interaction in the evolution of the perturbations. As seen from the above equations, the interaction simply introduces an additional term in the perturbed Euler equations for both dark matter and dark energy, which is driven by their relative perturbed velocity. This new term closely resembles a standard Thomson scattering, hence the name covariantised dark Thomson-like interaction because \eqref{eq:defQ} can be interpreted as the covariantization of this interaction. However, the key distinction between these interactions lies in the scales at which they operate. For the covariantised dark Thomson-like scattering to be effective, peculiar velocities between the interacting components must be present. As a result, this interaction is efficient only on small scales, where such peculiar velocities emerge, whereas on large scales the interaction does not have any effect because all components share the same rest frame. This is the same for a standard Thomson scattering. Regarding time-scales however, the interaction will play a relevant role whenever the condition $\Gamma \gtrsim \mathcal{H}$ is satisfied. For a constant coupling $\alpha$, as the one considered in this work, the interaction rate $\Gamma$ grows with the expansion as $a^4$ so the cosmological evolution naturally evolves from a weak coupling epoch to a strong coupling epoch, which means that the effects of the interaction naturally become relevant at late times. For a standard Thomson scattering, the effective interaction rate decays with the expansion and this leads to the opposite situation, namely: the cosmological evolution naturally leads to exiting a strong coupling phase so that the effects are less relevant at late times. This crucially different evolution in both scenarios explains why the covariantised elastic model described by equations \eqref{eq:deltaDM}, \eqref{eq:deltaDE}, \eqref{eq:thetaDM} and \eqref{eq:thetaDE} provides a relatively better performance than the pure Thomson scattering models. In particular, it is better suited for modifying the late time structure formation while leaving the core of the growth of structures at high redshift unaffected. The covariantised elastic model has been extensively studied in previous works \cite{Asghari:2019qld,Figueruelo:2021elm,BeltranJimenez:2021wbq,Cardona:2022mdq,BeltranJimenez:2022irm,BeltranJimenez:2024lml,Jimenez:2024lmm}, but only on linear cosmological scales. The goal of this work is to go beyond the linear regime and explore the non-linear scales. 
\begin{figure*}[t]
\centering
\includegraphics[width=18cm]{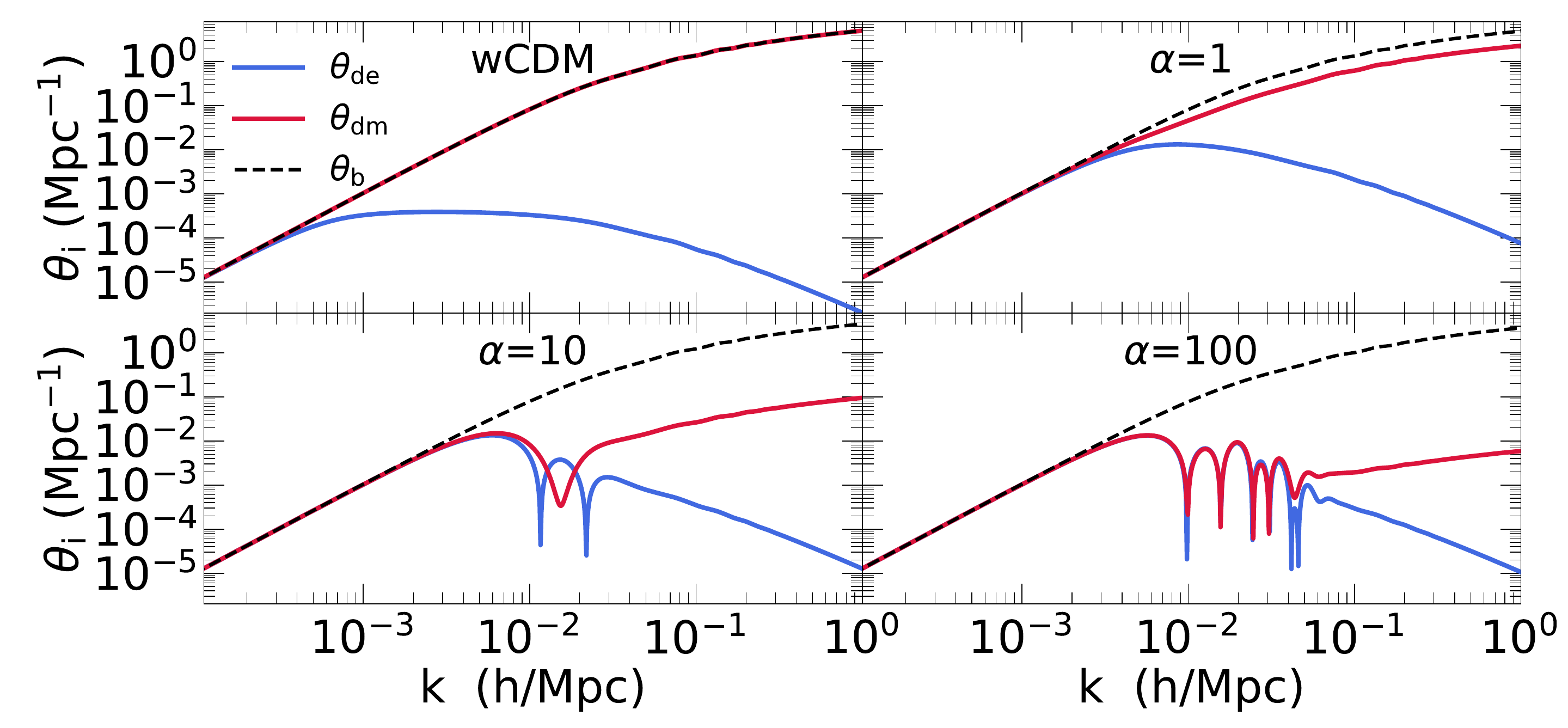}
\caption{Velocity divergence $\theta$ today of dark energy (blue), dark matter (red) and the non-interacting baryons (black) for different values of the coupling parameter $\alpha$. For the case analysed here with coupling parameter $\alpha=1$, we can corroborate that the velocity of dark energy is much smaller than those of baryons and dark matter, thus justifying  approximation done, $v_{\rm dm}\gg v_{\rm de}$ and $v_{\rm b}\gg v_{\rm de}$, of neglecting dark energy velocities for the scales relevant for our simulations.}
\label{fig:cov_dedm_DAO}
\end{figure*}

\section{Non-linear regime prescription}
\label{sec:non-linear}

\begin{figure*}[t]
\centering
\includegraphics[width=8.62cm]{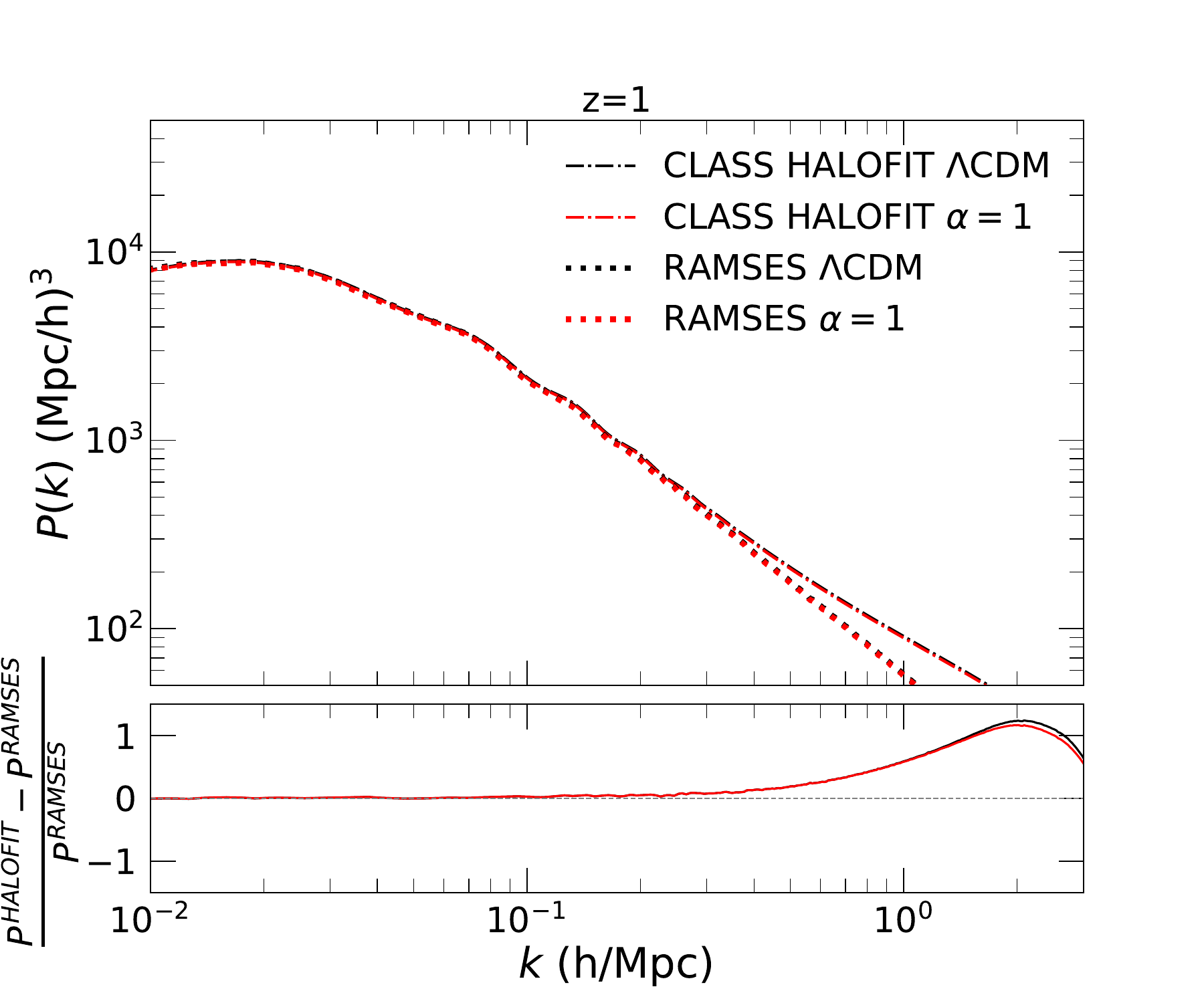}
\includegraphics[width=8.62cm]{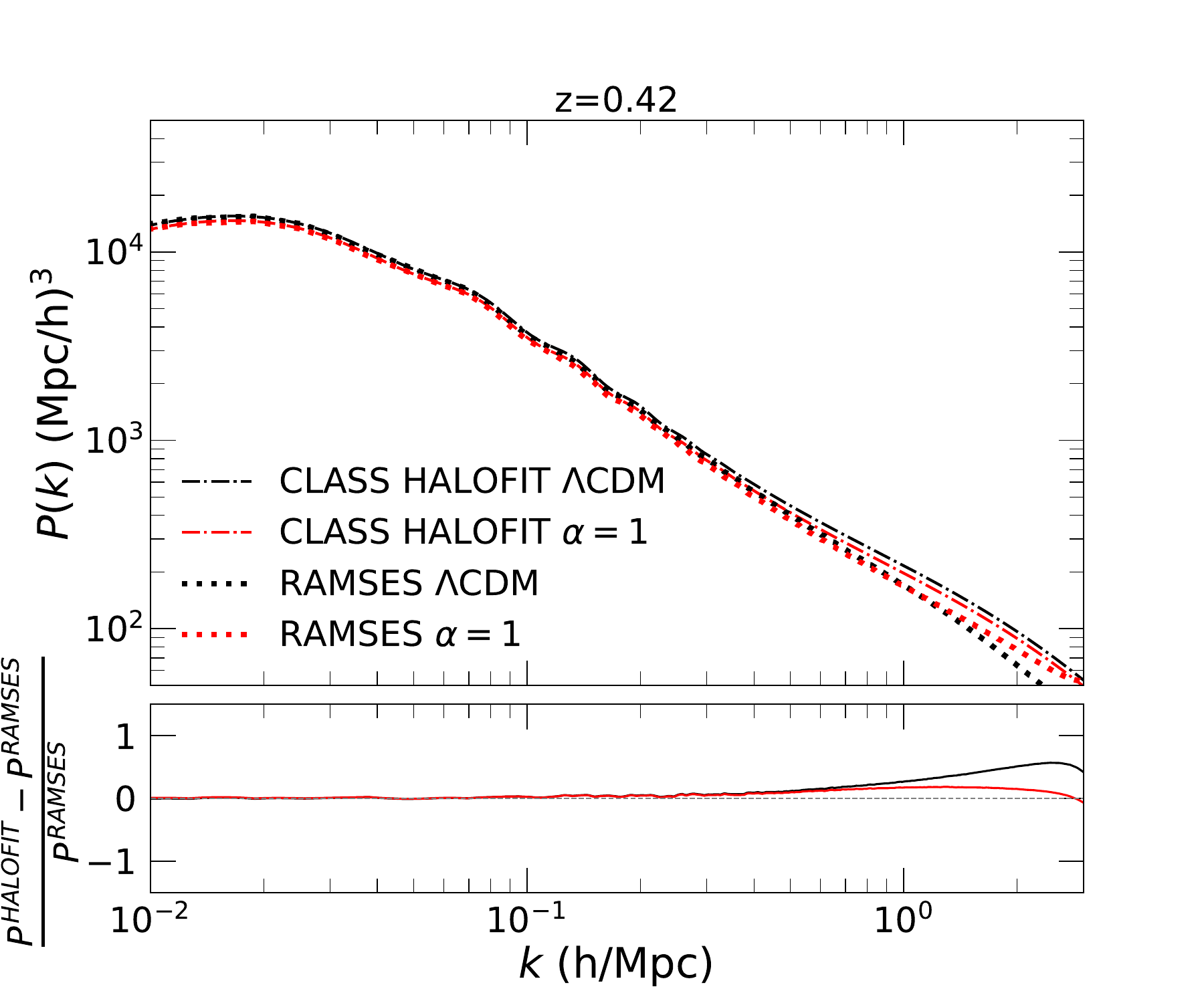}
\includegraphics[width=8.62cm]{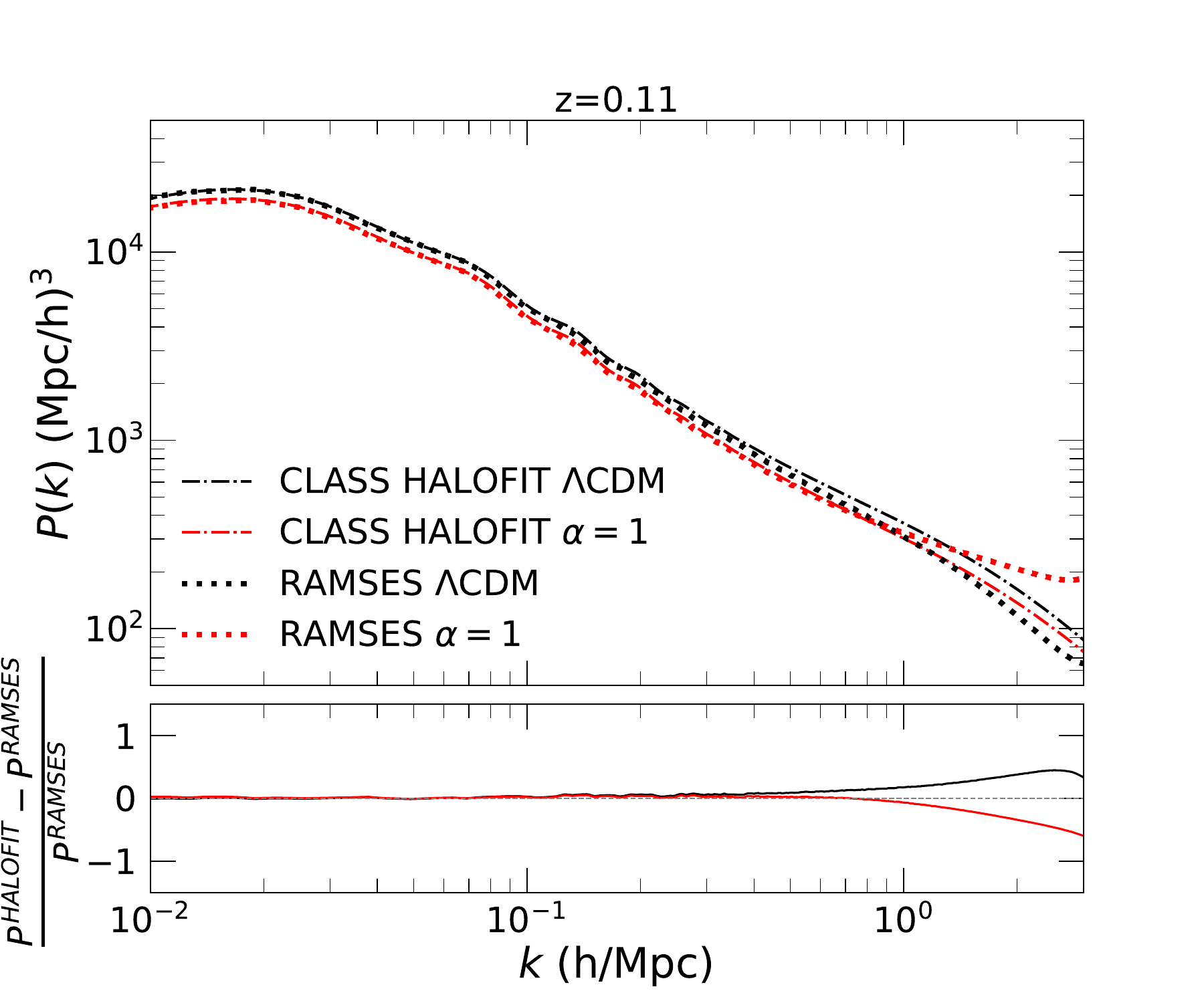}
\includegraphics[width=8.62cm]{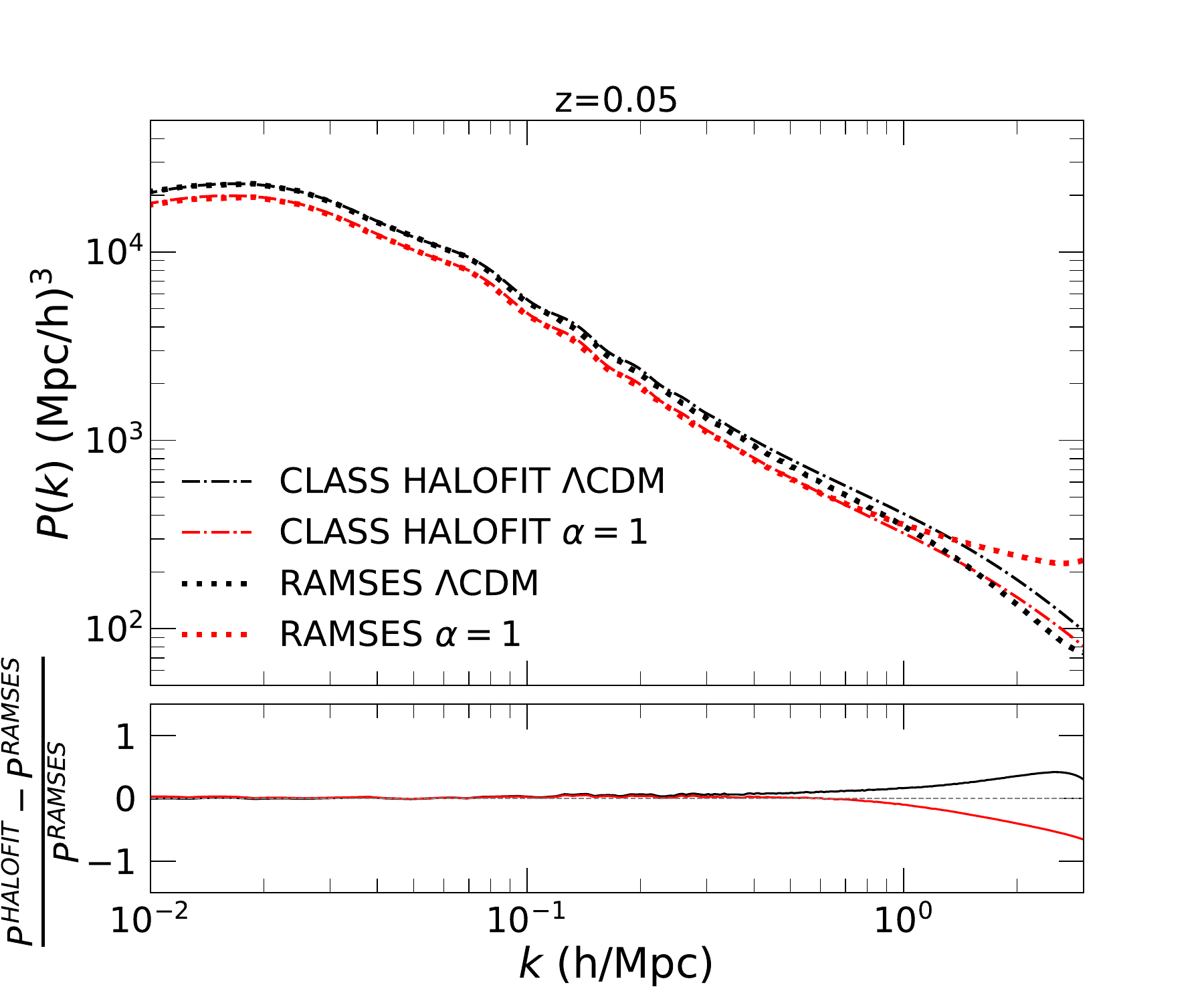}
\caption{The matter power spectrum for different redshifts computed from the modified version of \texttt{CLASS} with \texttt{HALOFIT} and from \texttt{RAMSES} (upper panels) and the ratio $P(k)^\texttt{HALOFIT}/P(k)^\texttt{RAMSES}-1$ in order to show to which precision \texttt{HALOFIT} can predict the non-linear scales (lower panels). 
}
\label{fig:nbody_pk}
\end{figure*}
\begin{figure*}[t]
\centering
\includegraphics[width=15cm]{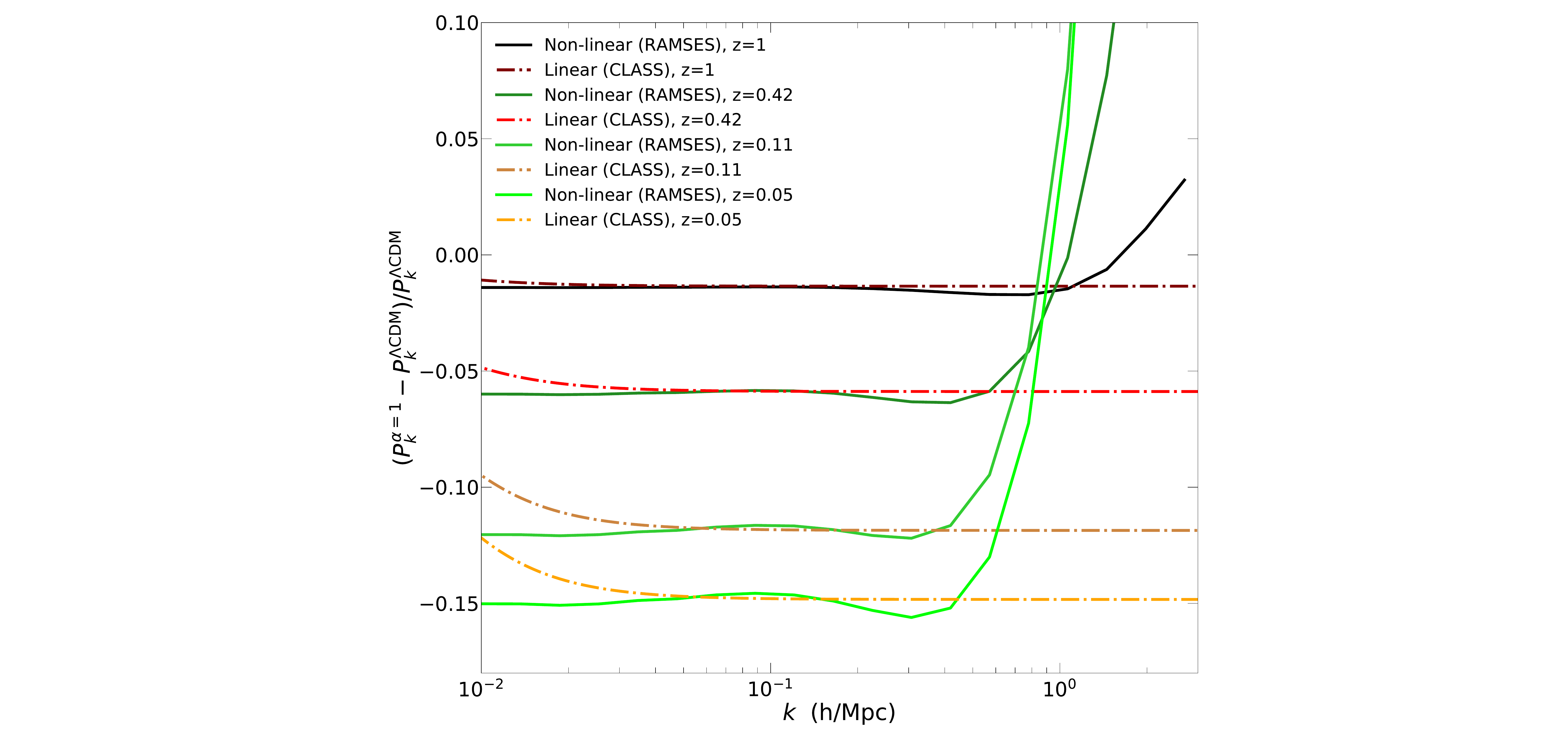}
\caption{ Relative difference for the matter power spectra between the covariantised dark Thomson-like interacting model and the $\Lambda$CDM model at different redshifts computed from the simulation output using all the particles (solid lines) and using the linear solver CLASS (dash-dotted lines).}
\label{fig:nbody_pk_relative}
\end{figure*}

To investigate the non-linear regime of the previously discussed elastic interacting model, we have implemented the corresponding  modifications into the $N$-body cosmological code \texttt{RAMSES}~\cite{Teyssier:2001cp} as we explain in the following.  Consider a simulation involving $N$ particles, where the $i$-th particle, which represents either dark matter or baryons, is subject to the gravitational force exerted by the remaining $(N-1)$-particles. In the Newtonian limit, within a cosmological background determined by the Hubble function $H$, the governing equations for the dynamics of the $i$-th particle can be expressed as follows:
\begin{eqnarray}
    \frac{{\rm d}^2 \Vec{x}_i}{{\rm d} t^2}+2H(t)\frac{{\rm d} \Vec{x}_i}{{\rm d} t}&=&- \frac{1}{a^2}\Vec{\nabla} \Phi(t,\Vec{x}_i)\;, \label{eq:master_v} \\
    \nabla^2 \Phi(t,\Vec{x})&=&4\pi G a^2 \overline{\rho}\delta(t,\Vec{x})\;,
    \label{eq:master_phi}
\end{eqnarray}
where $\Vec{x}_i$ are the comoving coordinates of $i$-th particle, $\Phi$ is the  gravitational potential, $H(t)$ is the Hubble function which encodes the cosmological model, while $\overline{\rho}$ and $\delta$ are the total energy density and the density contrast respectively. The first equation computes the acceleration of the $i$-th particle while the second one is the well-known Poisson equation. In the elastic interacting model, the Poisson equation remains unmodified, while the cosmological model encoded into the Hubble function is the $w$CDM model, consistent with the requirements of the model as previously explained. However, the acceleration of the particle is influenced not only by the expansion of the Universe and gravitational forces, but also by momentum transfer. Therefore, a new term will appear in equation~\eqref{eq:master_v} accounting for the dark energy drag.  From equation~\eqref{eq:thetaDM}, we can derive the corresponding acceleration equation for the covariantised dark Thomson-like scattering that now reads as
\begin{eqnarray}
\centering
\dot{\vec{v}}_i=-H\vec{v}_i-\frac{1}{a}\nabla\Phi-\alpha \frac{H_0}{\Omega_{{\rm dm}}(t)}\vec{v}_{i}\;,
\label{eq:master_Nbody_eq_alpha}
\end{eqnarray}
where we have rewritten the equation in terms of the velocities $\Vec{v}_i \equiv a\dot{\vec{x}}_i$ instead of the comoving coordinates $\Vec{x}_i$. In order to obtain the previous equation, we have performed the approximation $v_{\rm dm}\gg v_{\rm de}$ and $v_{\rm b}\gg v_{\rm de}$, implying that, on the relevant scales, the velocity of dark energy is negligible compared to that of matter.  N-body simulations typically focus on self-gravitating systems composed of a large number of particles and  generally do not account for a dark energy fluid, other than through its contribution to the Hubble function $H(z)$. However, it is important to emphasise that the focus here is on very small scales. In the standard scenario, on these non-linear scales, the velocity of dark energy is negligible, as gravitational forces dominate the dynamics. With the introduction of the coupling, however, this assumption may no longer hold. Nevertheless, as shown in Figure~\ref{fig:cov_dedm_DAO}, we demonstrate that when the interaction is efficient, the contribution of dark energy velocity to the new term in the Euler equation will remain negligible, as long as reasonable values for the coupling parameter $\alpha$ are considered. By reasonable we mean the values suggested by data in the previous MCMC analyses we have performed, from which we had obtained that $\alpha \sim\mathcal{O}(1)$ as seen in \cite{Figueruelo:2021elm}. This approximation will, however, produce inaccurate results when looking at larger scales where it no longer holds since we resort to the realms of the Cosmological Principle and, thus, the velocities of all components are of similar magnitude. To ensure proper distinction between the particles subjected to the interaction (dark matter) and those that are not (baryons), we introduced a "family" category to the particles of the simulation. Each particle is labelled either as dark matter, thereby experiencing the additional term in equation~\eqref{eq:master_Nbody_eq_alpha},  or as baryons, which do not experience the momentum transfer term. To ensure consistency, the relative number of particles assigned to each family follows the cosmology used, in particular matching the proportions given by the values of $\Omega_{\rm b}$ and $\Omega_{\rm dm}$ provided below.

\begin{figure*}[t]
\centering
\includegraphics[width=8.62cm]{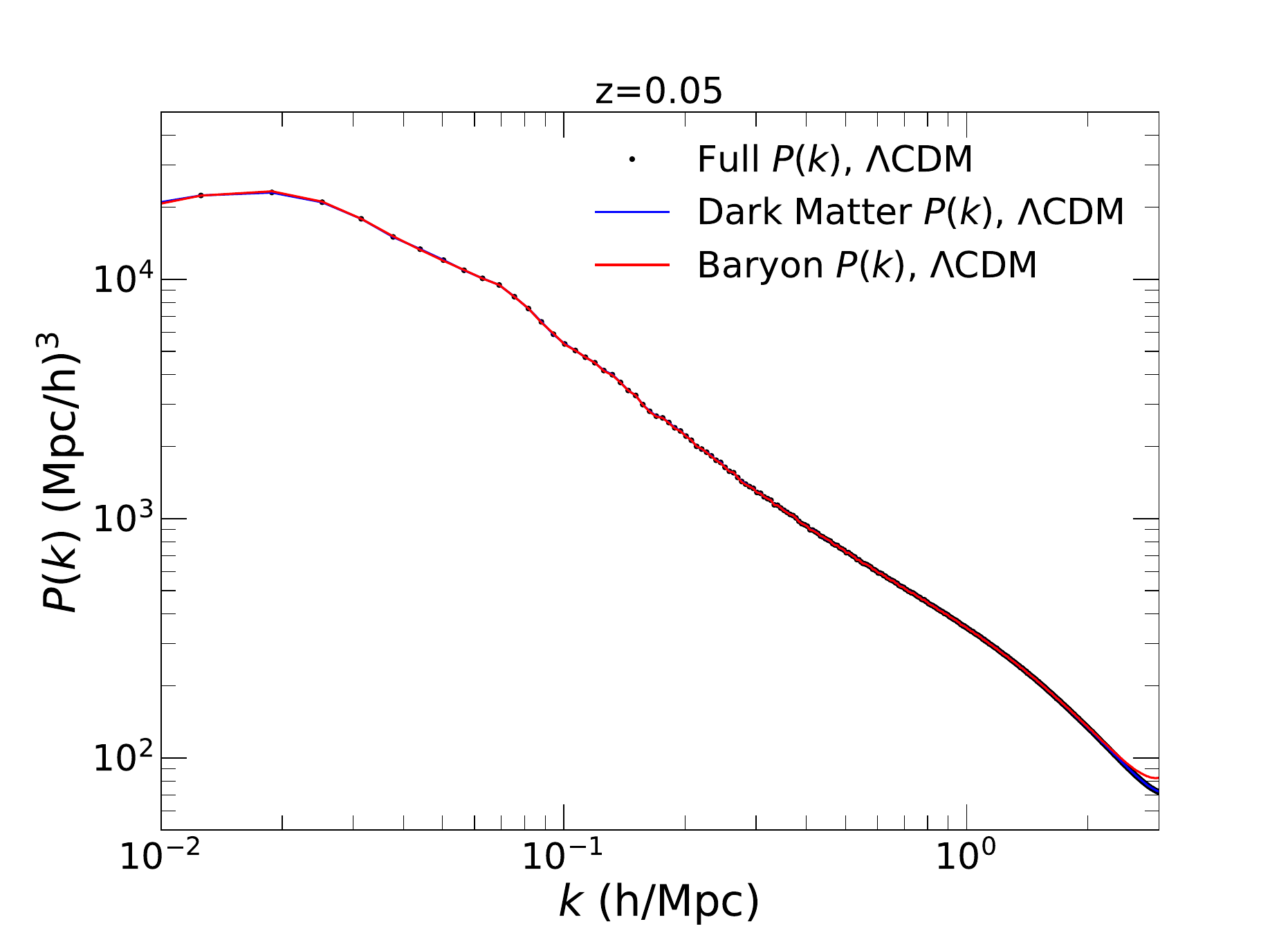}
\includegraphics[width=8.62cm]{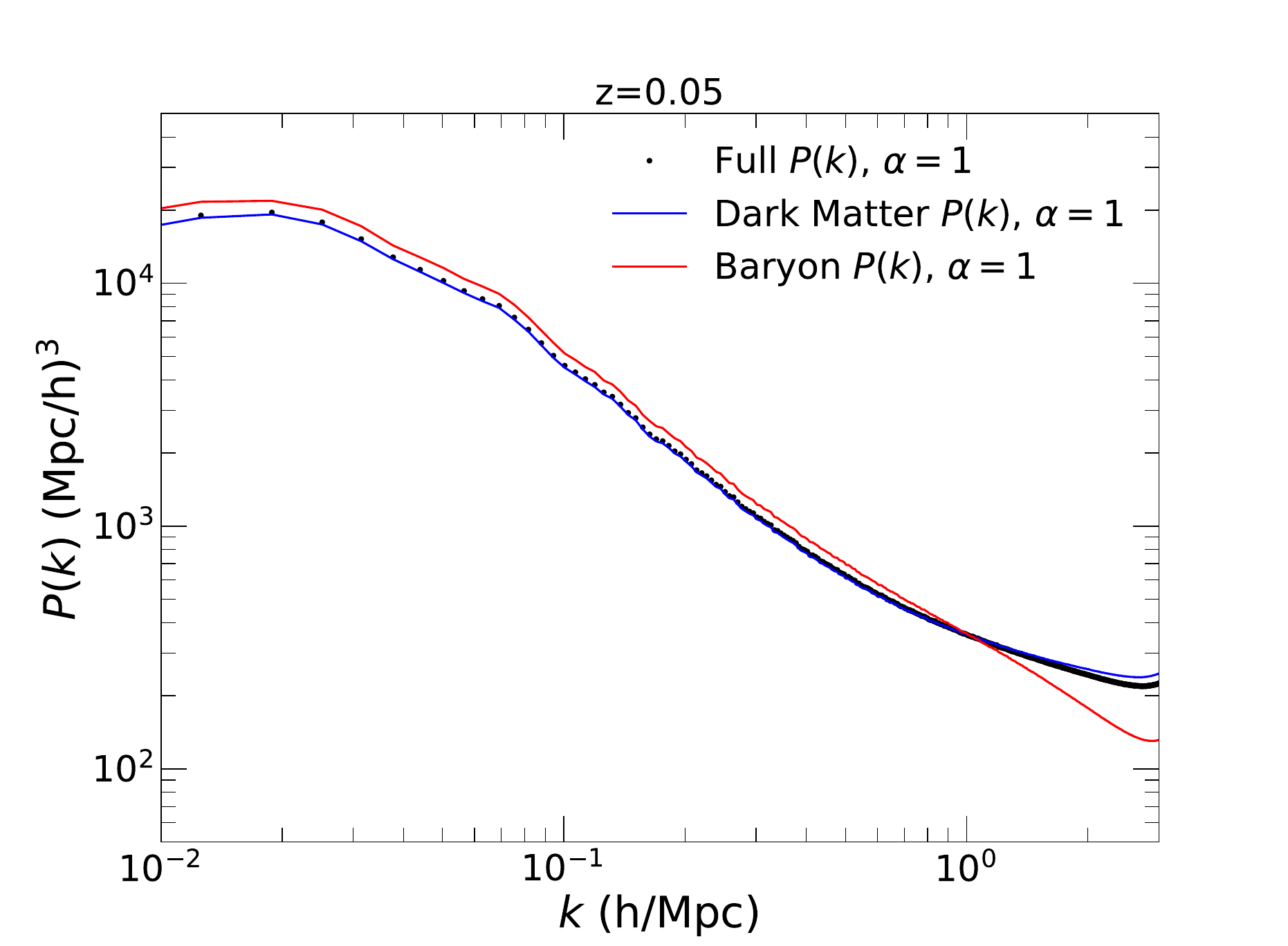}
\caption{The matter power spectrum for $\Lambda$CDM model (left) and for the covariantised dark Thomson-like interacting model at $z=0.05$ computed from the simulation output using all the particles (dot-points), only dark matter particles (blue line) and only baryons particles (red line).}
\label{fig:nbody_pk_separated}
\end{figure*}

\section{Results from the simulations}
\label{sec:results}
In this work, we consider a cosmology described by the following parameters. The Hubble parameter is fixed to be $H_0=67.7\;{\rm km/s/Mpc}$, the today's relative density parameters as $\Omega_{\rm b}=0.045$, $\Omega_{\rm dm} =0.269$, while for the scalar perturbations we have $A_{\rm s}=2.1\times 10^{-9}$ and $n_{\rm s}=0.968$, and dark energy is described by a constant equation of state as $w=-0.98$. The initial conditions were obtained from the \texttt{MUSIC2-monofonIC} code \cite{2020ascl.soft08024H,2021MNRAS.503..426H} with the quoted choice of parameters. The analyses are performed for a box with $L=10^3\;\text{Mpc}/h$ populated with $N=512^3$ particles. We have performed two simulations:
\begin{itemize}
    \item $\Lambda$CDM model with the previous choice of cosmological parameters.
    \item Covariantised dark Thomson-like scattering with coupling parameter $\alpha=1$ and the same choice of cosmological parameters. The initial conditions for this simulation are the same as for the $\Lambda$CDM one, which is justified because the interaction is irrelevant at the time when we set the initial conditions. Additionally, only dark matter particles evolve according to equation~\eqref{eq:master_Nbody_eq_alpha}, while baryon particles follow the standard evolution given by equation~\eqref{eq:master_v}.
\end{itemize}

Let us now proceed to analysing the results from our simulations, where we will compare to $\Lambda$CDM to study the effects of the interaction in the considered non-linear scales.

\subsection{Matter Power Spectrum and check with linear results}
In Figure~\ref{fig:nbody_pk}, we present the evolution of the matter power spectrum across various redshifts, while in Figure~\ref{fig:nbody_pk_relative} we display the same redshift evolution but now for the relative difference between the matter power spectrum in the interacting case and the reference $\Lambda$CDM model. With the exception of very large scales $k\lesssim (aH)_0$, where our numerical implementation is expected to produce unreliable results\footnote{On very large scales, the velocity of dark energy becomes comparable to that of matter so our approximation $v_{\rm dm} \gg v_{\rm de}$ and $v_{\rm b} \gg v_{\rm de}$ breaks down. Since we are interested in small (non-linear) scales, this is not relevant for our purposes here. However, this should be refined for simulations including e.g. relativistic effects and/or if near horizon scales effects are to be analysed.}, the primary effects of the interaction are predominantly observed on non-linear scales and only become significant at relatively late cosmic times. This behaviour is consistent with the results obtained from linear perturbation theory.\\
At redshift $z=1$, shown in the first panel, no noticeable deviation is observed between the standard cosmological model and the interacting scenario. Although already known from previous works (see \cite{Asghari:2019qld,BeltranJimenez:2021wbq,Cardona:2022mdq,BeltranJimenez:2024lml}) this supports our use of the same initial conditions for both $\Lambda$CDM and the interacting model simulations. However, as the interaction starts affecting the evolution when $\Gamma \gtrsim \mathcal{H}$, its influence first manifests on smaller scales, as apparent in the second panel at redshift $z=0.42$. The effect is a reduction of the clustering since the momentum transfer prevents dark matter from efficiently falling into the gravitational potential wells.  As cosmic evolution progresses and the interaction further intensifies, the range of affected scales expands towards larger scales. This increasing impact of the drag leads to a suppression of structure formation, which is reflected in a reduced clustering amplitude on a wider range of scales. Consequently, the matter power spectrum in the interacting scenario with $\alpha=1$ exhibits a systematically lower amplitude compared to the standard cosmological model on smaller scales. This trend is in full agreement with previous results derived in the linear regime as we can see comparing the results from the Boltzmann solver \texttt{CLASS} using the \texttt{HALOFIT} non-linear prescription~\cite{2011arXiv1104.2932L,2011JCAP...07..034B}. We can however see that \texttt{HALOFIT} tends to underestimate the suppression on small scales. This agrees with the expectation stated in \cite{Lague:2024sox} where the authors developed a halo-model approach to include quasi-linear scales for the elastic interacting model considered in this work, that they refer to as $w\Gamma$CDM. The authors rightfully recognize that \texttt{HALOFIT} is expected to underestimate the impact of the interaction in the non-linear regime and our full $N$-body  simulations confirm this expectation.

Finally, at highly non-linear scales, we observe an increase in the power spectrum indicating an enhancement of the dark matter clustering in the presence of the elastic interaction. This effect is opposite to the expected suppression of structures induced by the interaction by naively extrapolating the results of the linear scales. One may be tempted to ascribe this effect to a numerical issue related to the resolution of the simulations, but Figure~\ref{fig:nbody_pk_separated} clearly shows the effect is driven by dark matter and, thus, it is physical. A similar effect has also been observed in \cite{Baldi:2014ica,Baldi:2016zom} for the elastic scattering scenario between dark energy and dark matter introduced in \cite{Simpson:2010vh}. In those works, it is nicely explained how this enhancement in the clustering can be understood in terms of the non-linear virialisation process. On non-linear scales where dark matter particles start acquiring angular momentum, the interaction catalyses a drain of kinetic energy that modifies the virialisation process. Having less kinetic energy, the virialised structure shrinks with respect to the non-interacting case and this causes a more clustered matter distribution. This explanation will be later confirmed in section~\ref{subsec:halos_indivi} when we study the profile of the dark matter halos under the effect of the model, finding that indeed structures tend to shrink under the effects of the interaction.  The simulations in \cite{Baldi:2014ica} are for dark matter only while we have also included a non-interacting component as a proxy for baryons so we can see the effects on the baryons as well. This is shown in Figure~\ref{fig:nbody_pk_separated} where we can see how the enhanced clustering on non-linear scales is substantially more pronounced for the interacting DM, while baryons are only mildly sensitive to the effect because they only feel it through the modification to the gravitational potential. In the halo-model, based on the spherical collapse of \cite{Lague:2024sox}, it was also found an increase in the power spectrum on small scales that was attributed to mode-coupling to large-scale modes. While their results are expected to be valid only on quasi-linear scales, our results show that this trend extends to fully non-linear scales. Let us finally  notice that the enhancement of the power spectrum on non-linear scales has also been observed in a scenario with an elastic scattering between dark energy and baryons in the simulations run in \cite{Ferlito:2022mok}. The effect found in that work is however milder for two reasons, namely: baryons are less abundant than dark matter and the interaction rate from the elastic Thomson scattering reduces with the expansion.

\subsection{Halos: global picture}

\begin{figure}[H]
\centering
\includegraphics[width=8.50cm]{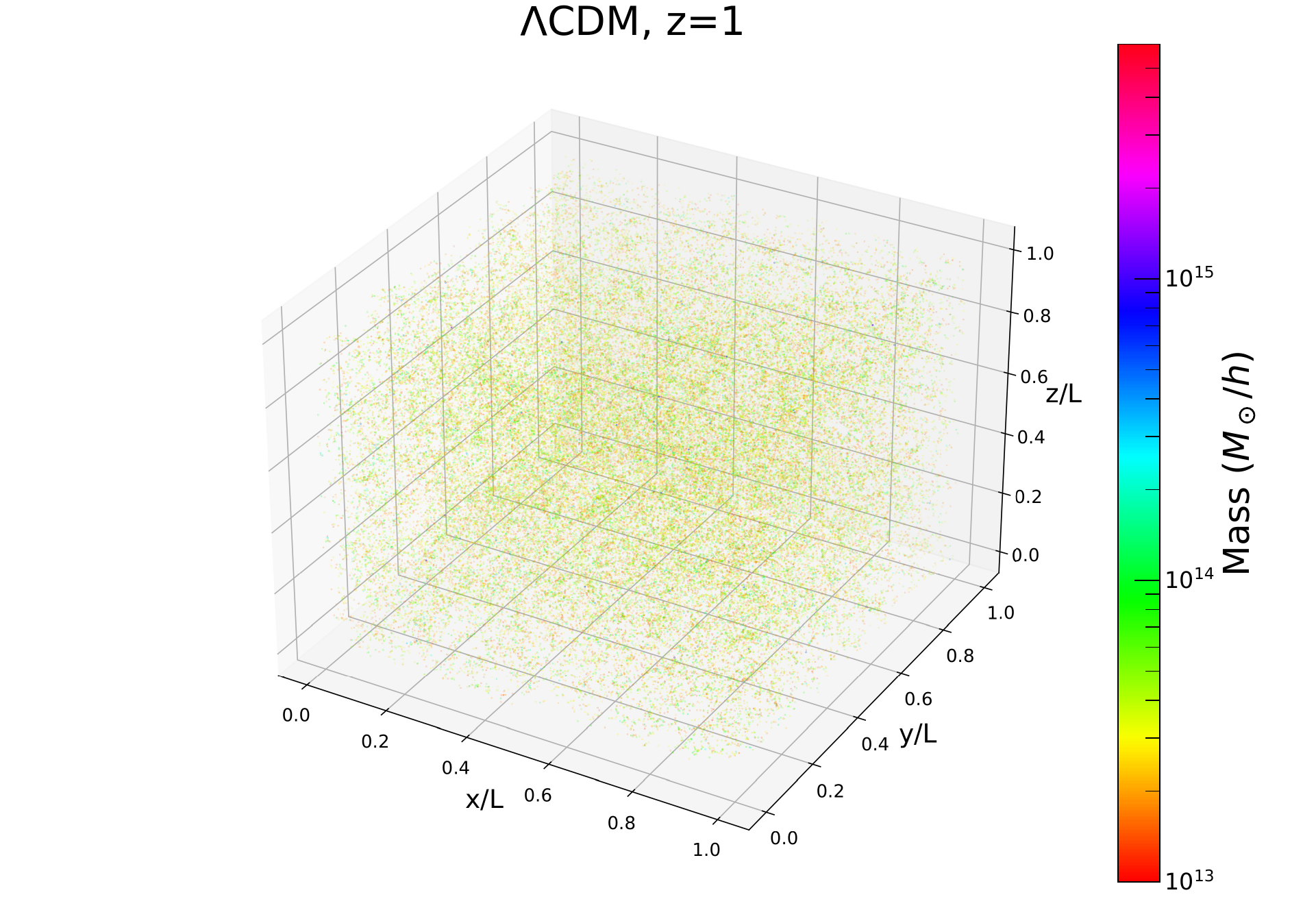}
\includegraphics[width=8.50cm]{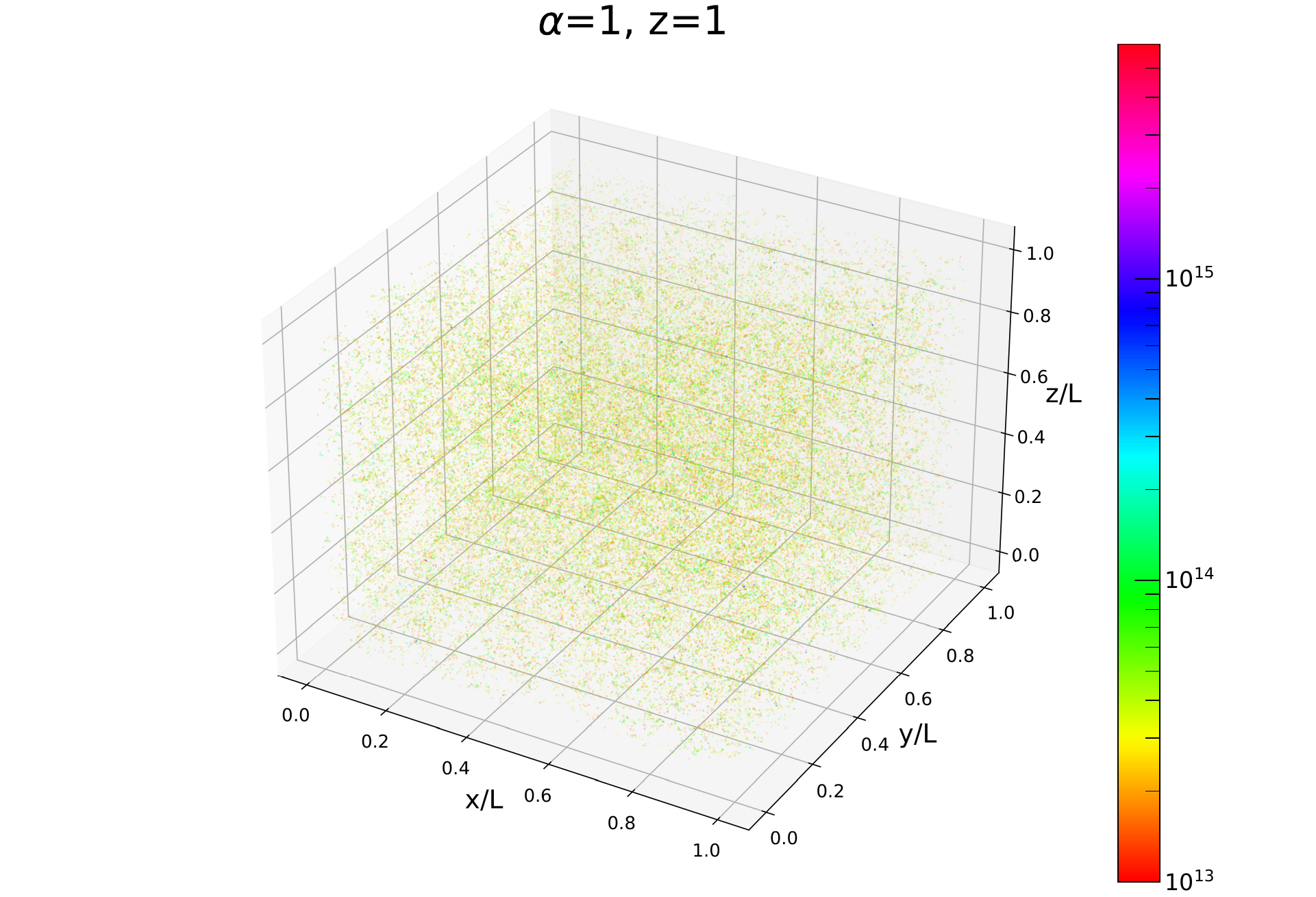}
\includegraphics[width=8.5cm]{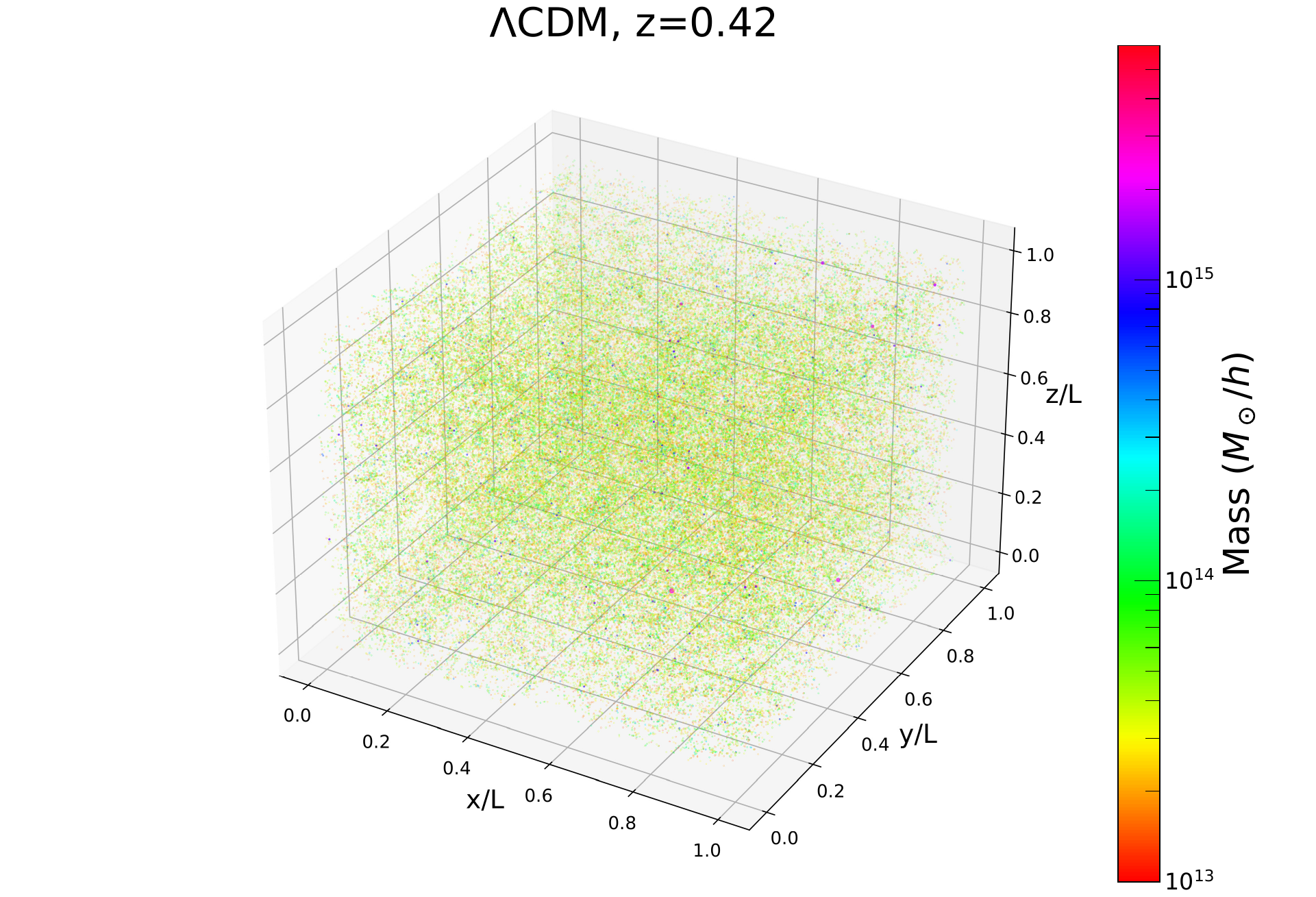}
\includegraphics[width=8.5cm]{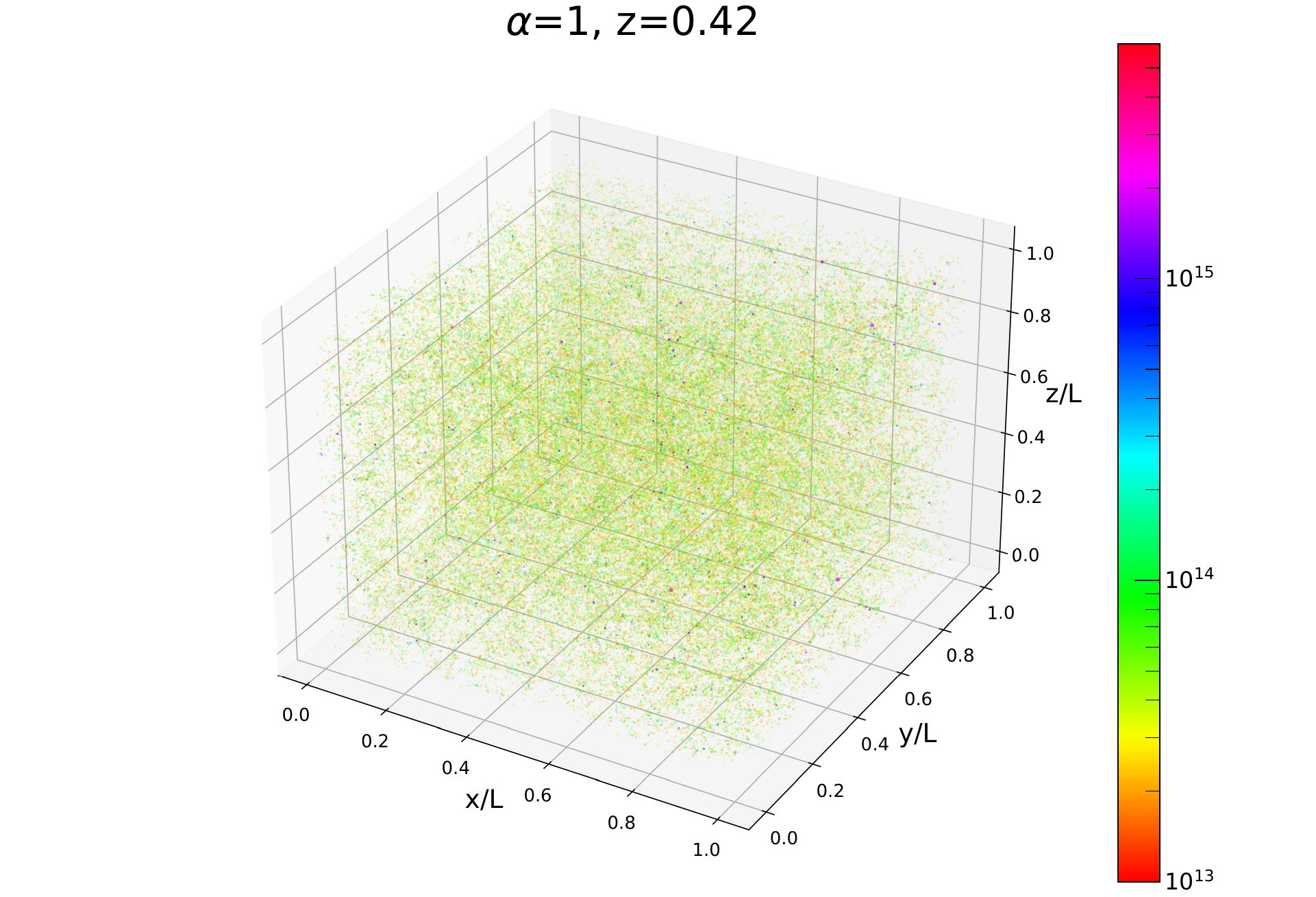}
\includegraphics[width=8.5cm]{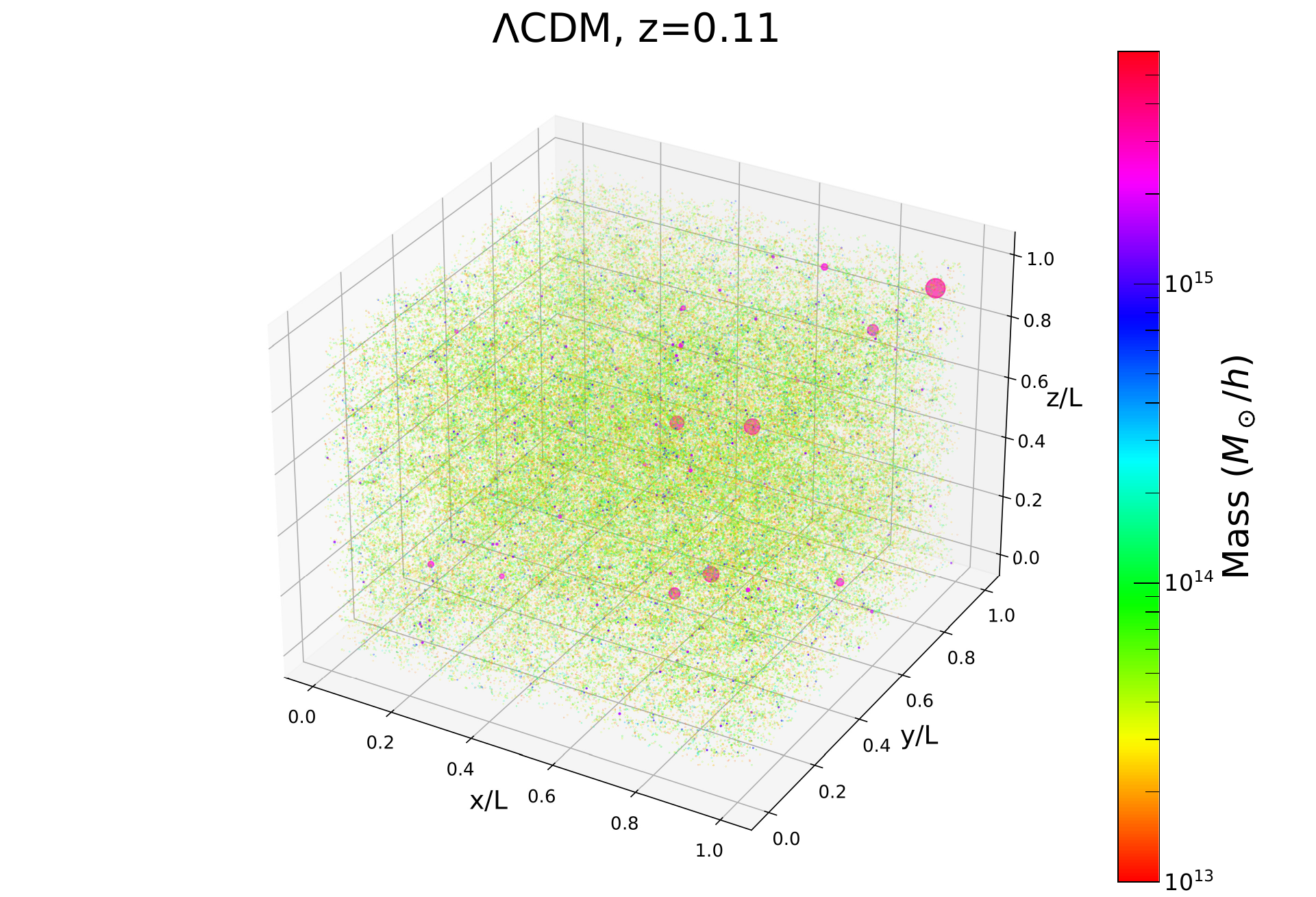}
\includegraphics[width=8.5cm]{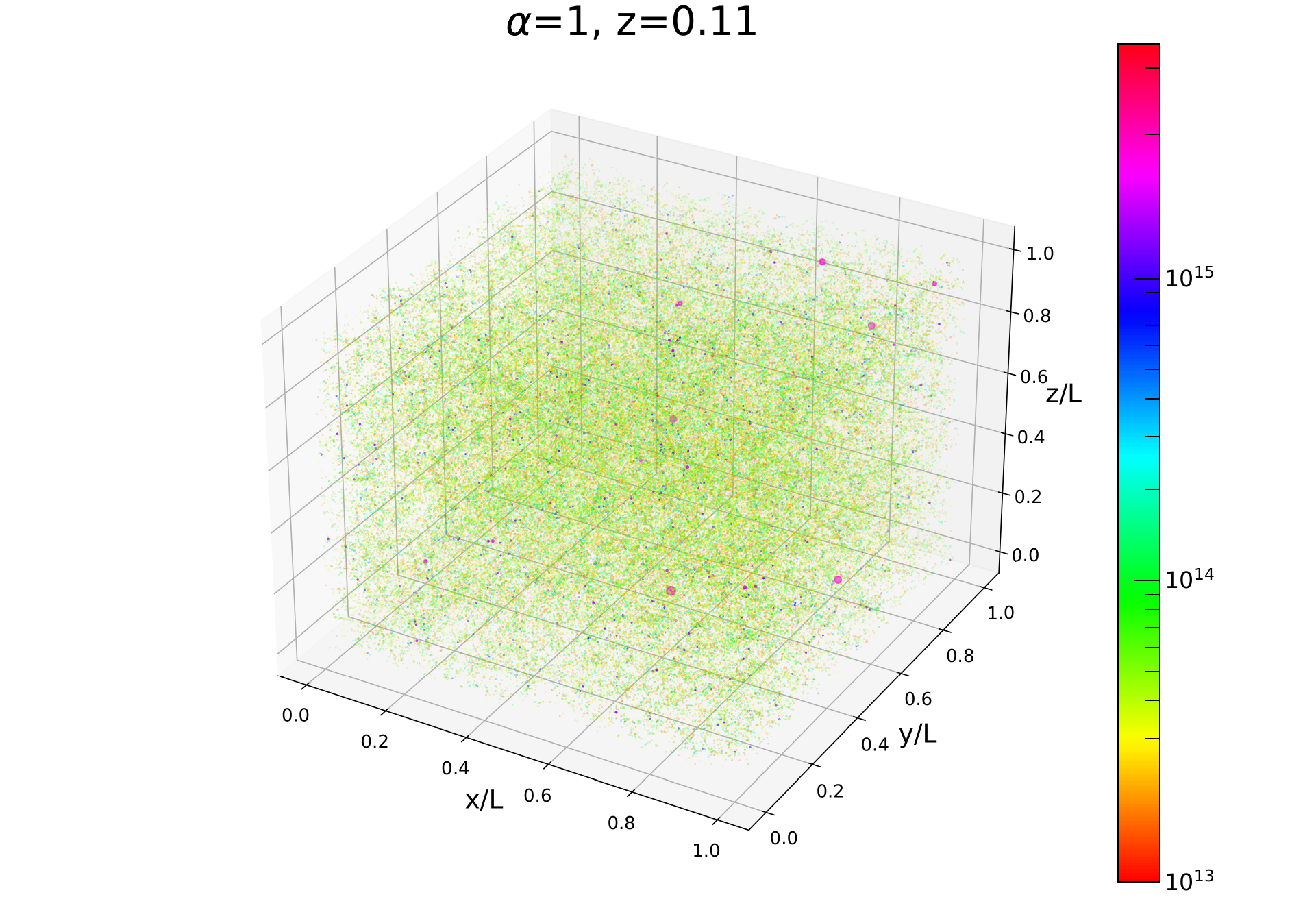}
\includegraphics[width=8.5cm]{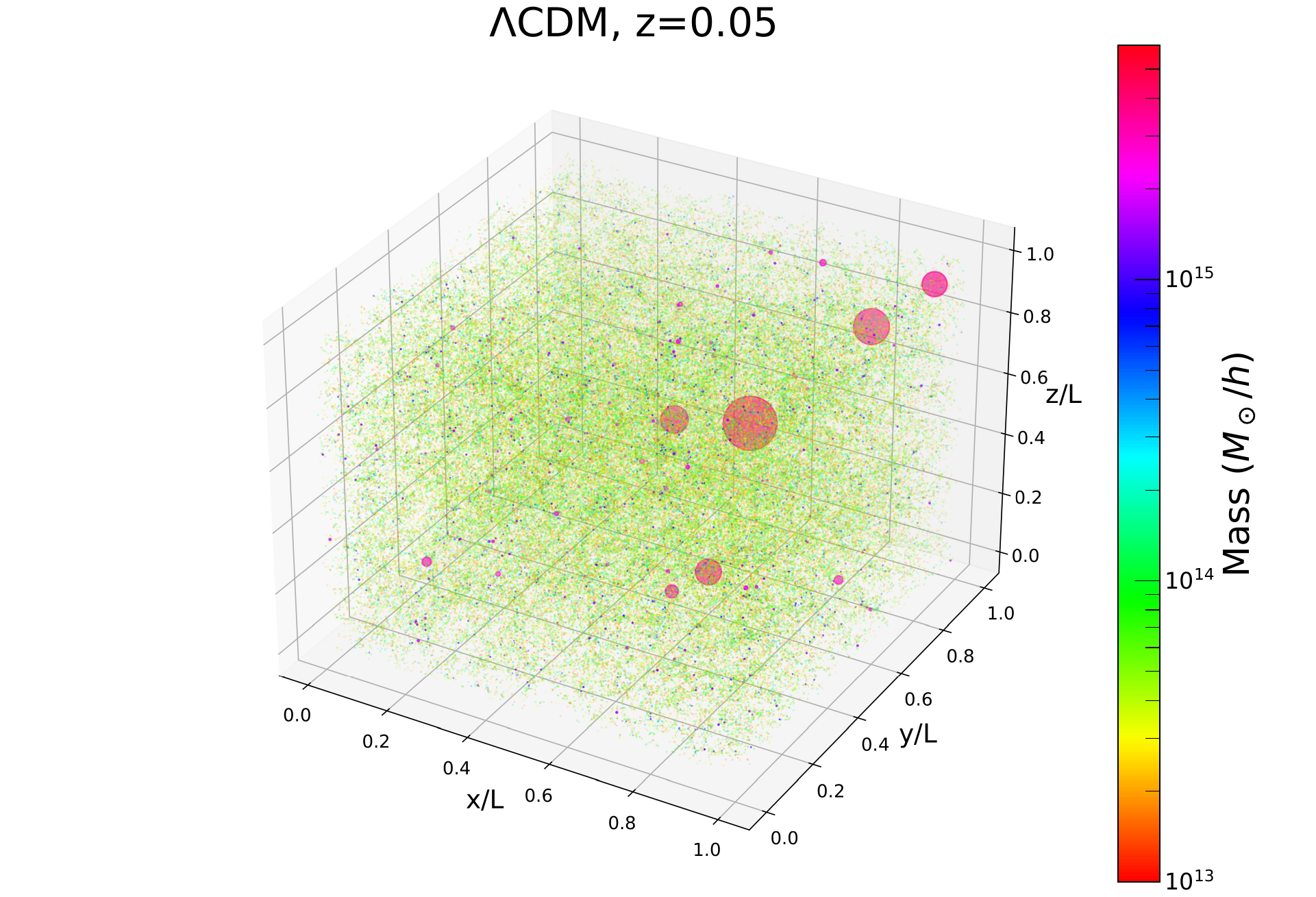}
\includegraphics[width=8.5cm]{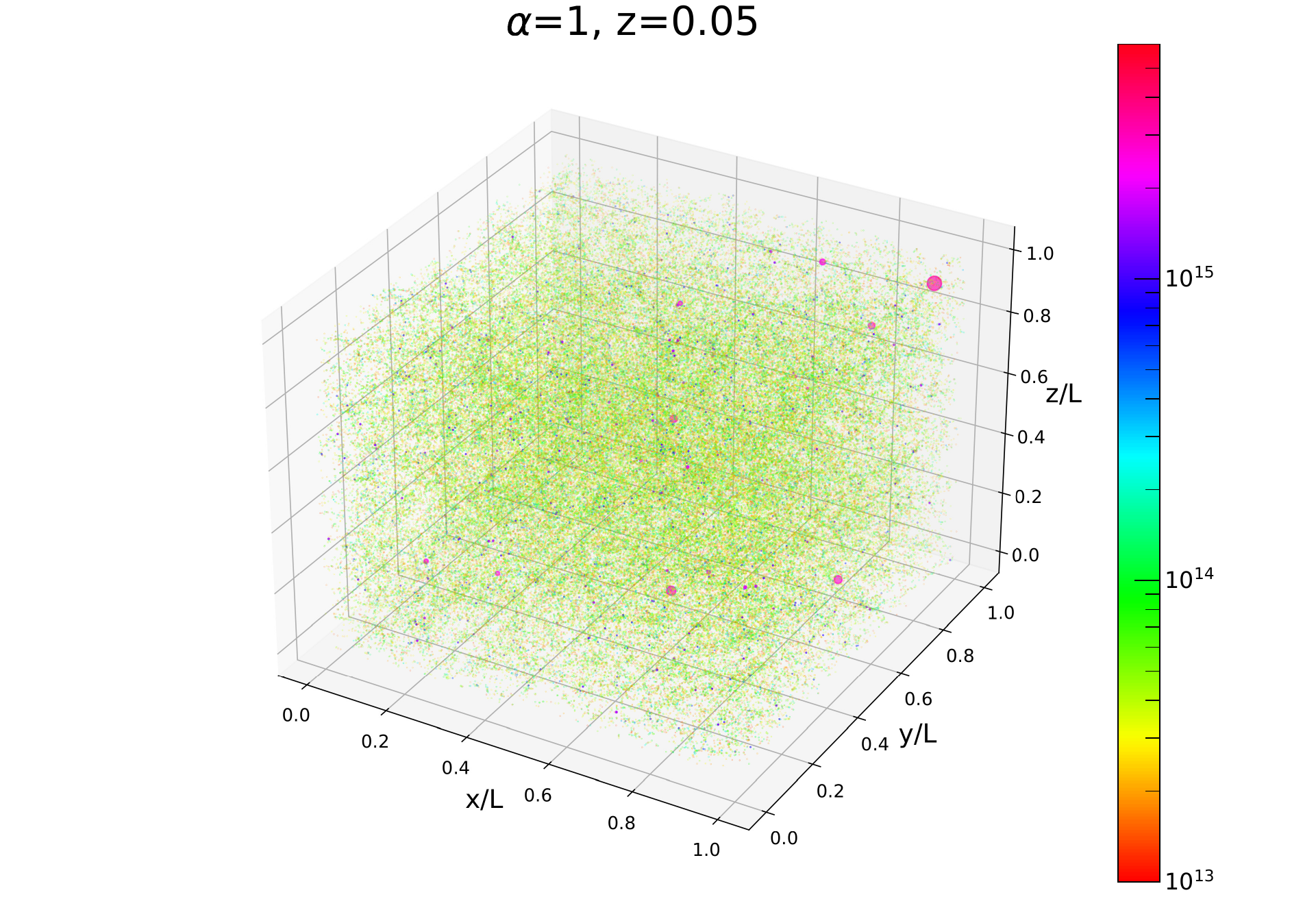}
\caption{Spatial distribution of the halos found by \texttt{MatchMaker} in the simulation box. Each dot is a halo whose colour and dot-size are proportional to the halo mass.}
\label{fig:nbody_distro_a}
\end{figure}

\begin{figure}[ht]
\centering
\includegraphics[width=8.62cm]{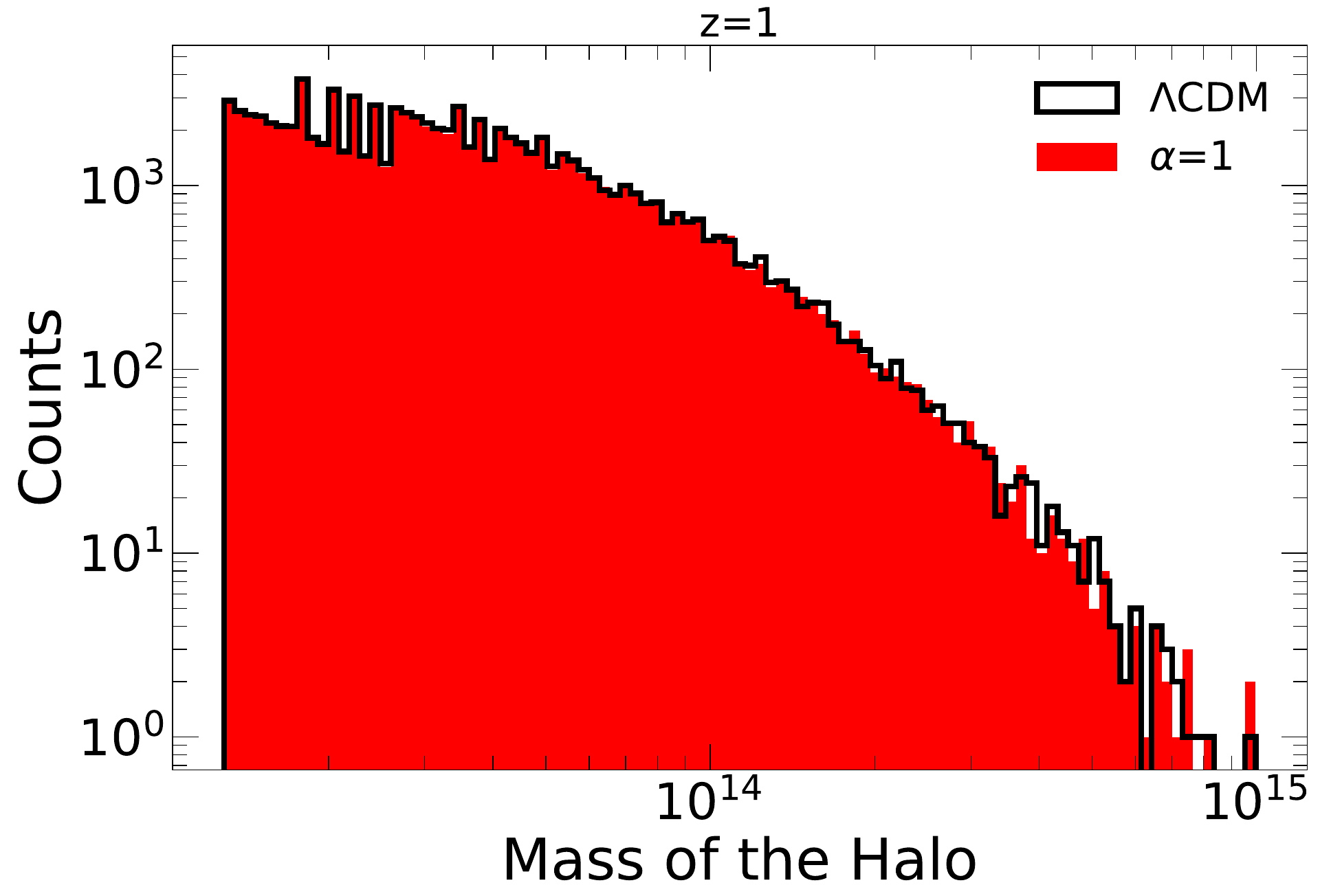}
\includegraphics[width=8.62cm]{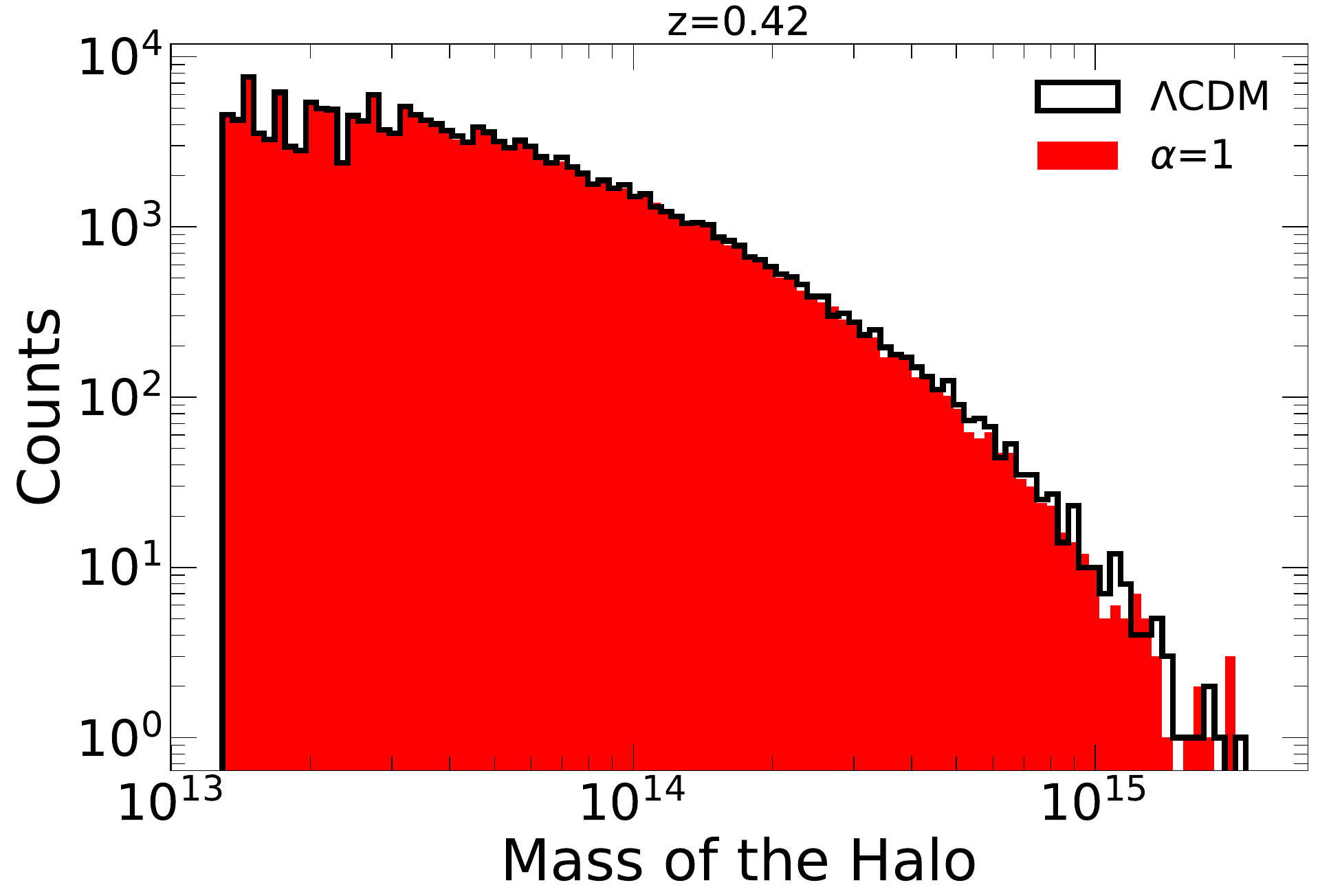}
\includegraphics[width=8.62cm]{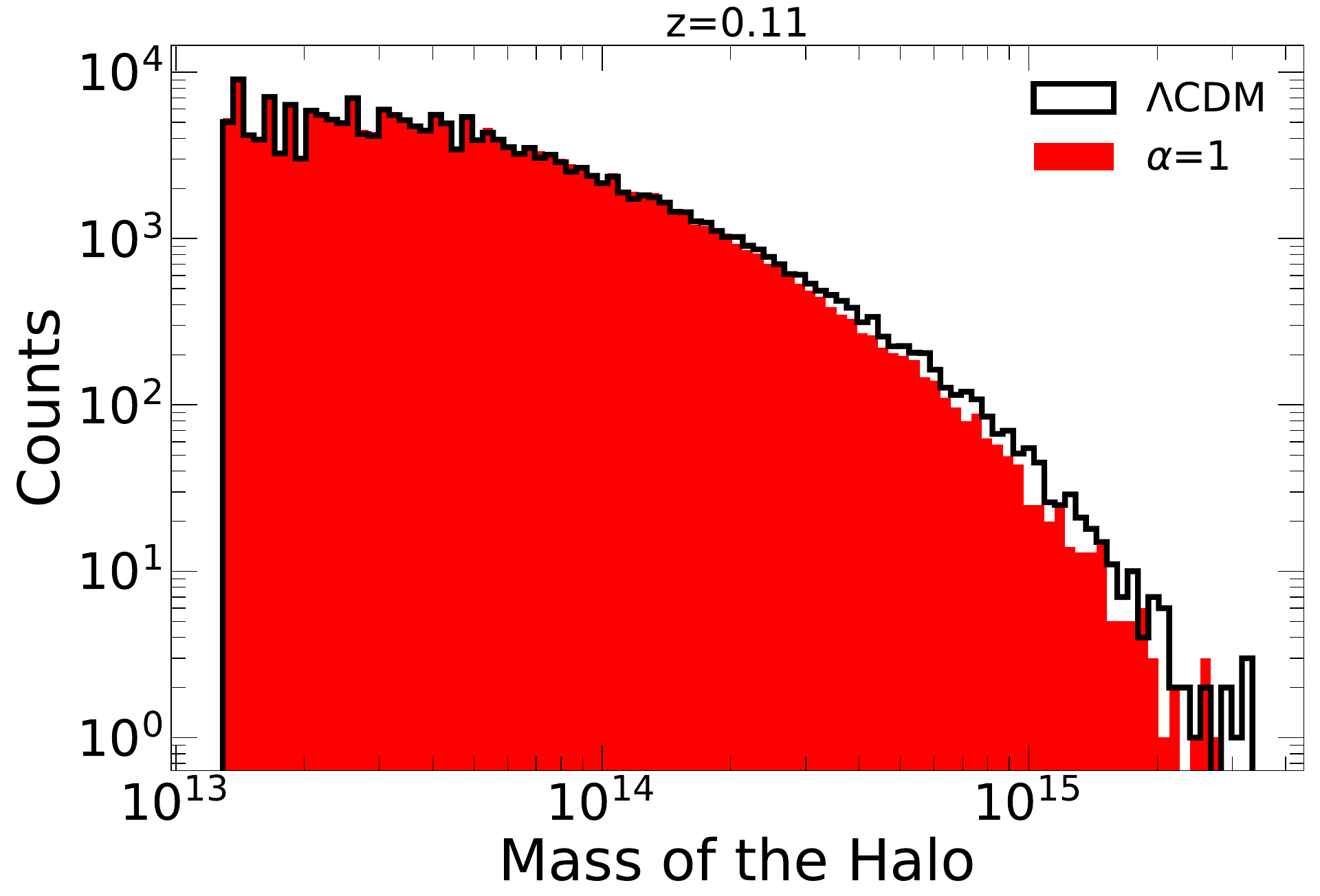}
\includegraphics[width=8.62cm]{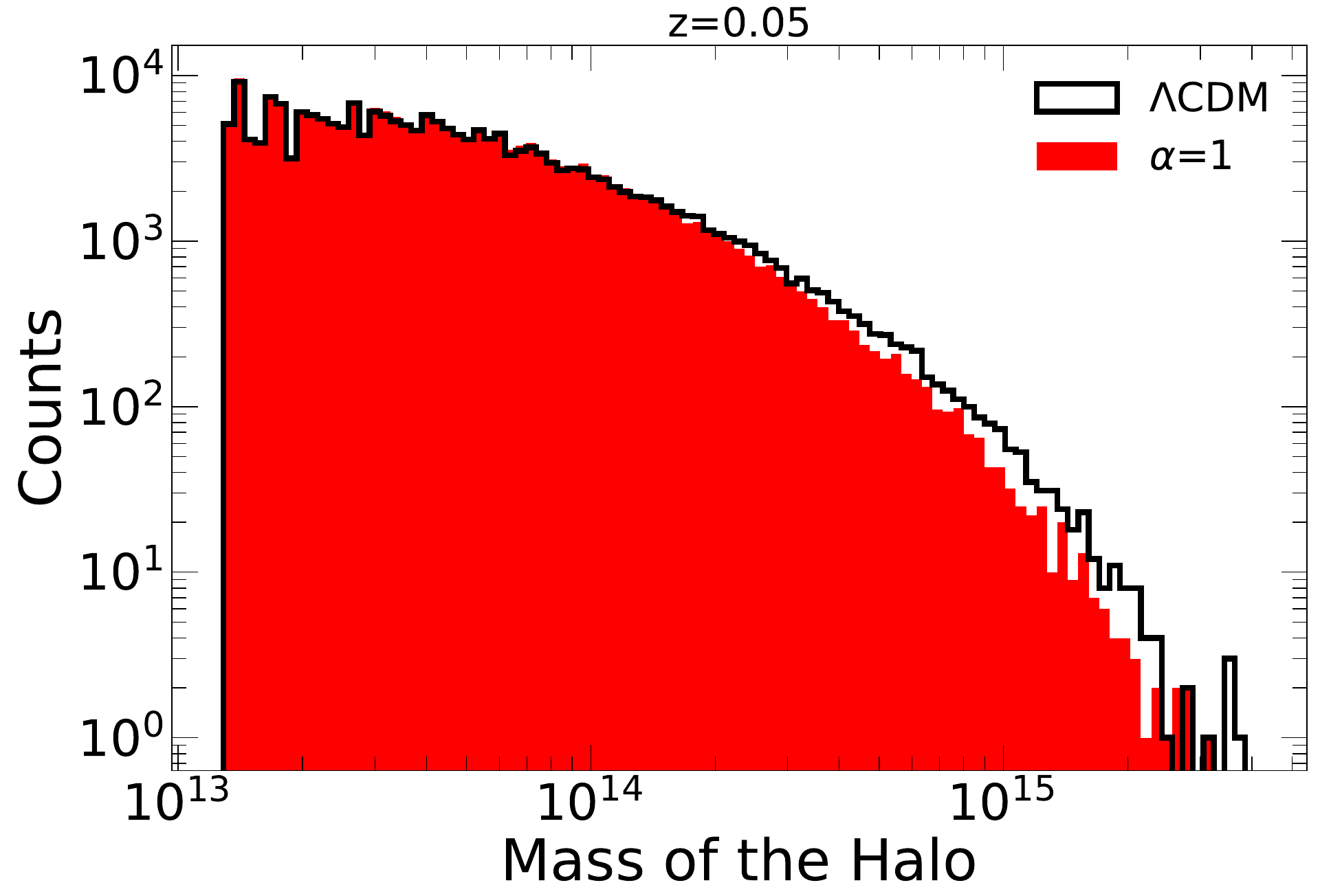}
\caption{Distribution of halos according to their mass in $M_{\odot}/h$ units for the standard model (black) and for the  covariantised dark Thomson-like interaction (red). }
\label{fig:nbody_part}
\end{figure}

We now turn our attention to how different halos form and are distributed. To analyse this, we use \texttt{MatchMaker}\footnote{MatchMaker is available at \href{https://github.com/damonge/MatchMaker}{https://github.com/damonge/MatchMaker}.}, a friend-of-friend halo finder, with the standard parameters, namely: a linking length of $b_{\rm fof}= 0.2$ in units of the mean inter-particle distance, and a minimum of $n_{\rm min}= 20$ particles per halo. Since we are focusing on the overall behaviour, we do not distinguish between dark matter and baryon particles when identifying halos, though it is important to note that the interaction does differentiate between them, as we will analyse later. 
With these considerations in mind, Figure~\ref{fig:nbody_distro_a} shows the spatial distribution of halos detected by \texttt{MatchMaker}. Each dot represents an individual halo, and the colour and size of the dot correspond to the mass of the halo, measured in units of $M_{\odot}/h$. Comparing the result from the $\Lambda$CDM and the interacting model simulations, we can already see how the interaction suppresses the formation of very massive halos while those of medium and small mass seem to be insensitive in terms of the amount of them that are formed. Considering halo formation is a hierarchical process in the CDM paradigm, we know that those very massive halos form only at very late times and from the merging of smaller halos. Therefore, as the momentum transfer for $\alpha=1$ happens only at low redshift, it is natural to infer why the most massive halos are the ones prevented from forming. In other words, the most massive halos that are about to form when the interaction becomes efficient, $\Gamma \gtrsim \mathcal{H}$, are the ones affected by the clustering deficit of the interacting model. On the other hand, from our analyses we can finally infer that the momentum transfer does not disrupt already formed structures but only halts the accretion process, preventing further growth of density perturbations.
To explore the formation of halos in more detail, we can perform a halo number count analysis. In Figure~\ref{fig:nbody_part}, we present the number of halos formed as a function of their mass. The differences between the $\Lambda$CDM model and the interacting scenario are not statistically significant for lower-mass halos. Again, those halos are typically already formed when the interaction becomes effective. Then, we can now quantitatively conclude the amount of   smaller and medium size halos formed is insensitive to the nature of this interaction. The scale in terms of mass  where the halos are affected is fixed by the same parameter that controls the strength of the interaction, $\alpha$, since it also controls the time scale at which the interaction becomes efficient. 
For larger halos than that scale, significant differences emerge. At early times, such as $z=1$, when the interaction has yet to take effect, no  differences are observed. This is expected, as both scenarios begin with the same initial conditions and follow similar evolutionary histories, leading to identical early structure formation in both simulated Universes. Once the interaction becomes efficient, however, the formation of very massive halos is suppressed in the momentum transfer simulation. This trend aligns with the patterns observed in the spatial distribution of halos. We can conclude that, in terms of amount of structure formation, only the most massive ones about to form at the onset of the interaction are going to be significantly modified. Furthermore, this modification manifests as a reduced probability of their formation, ultimately resulting in a Universe with fewer very massive structures.

\subsection{Halos: Individual picture}
\label{subsec:halos_indivi}

\begin{figure*}[t]
\centering
\includegraphics[width=8.62cm]{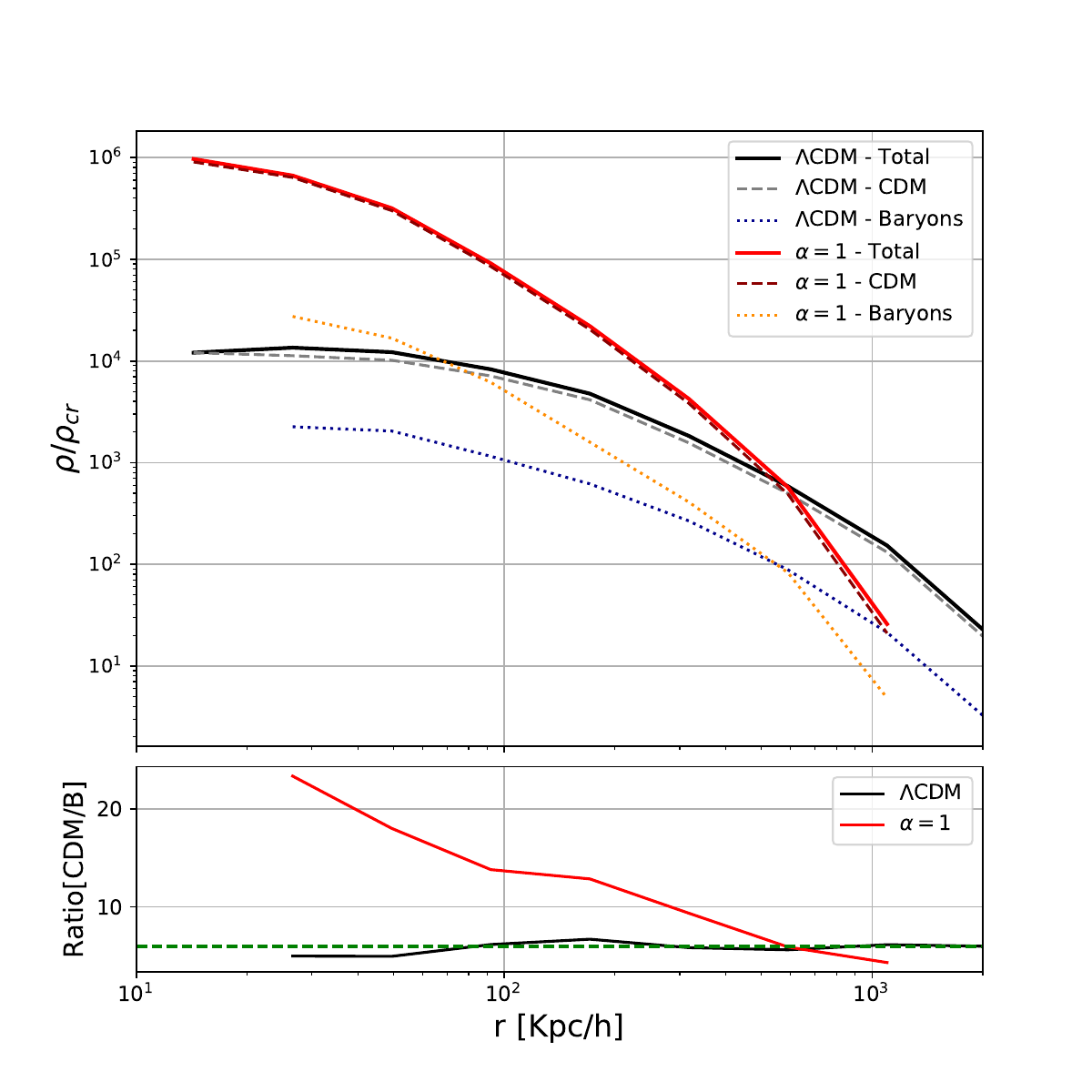}
\includegraphics[width=8.62cm]{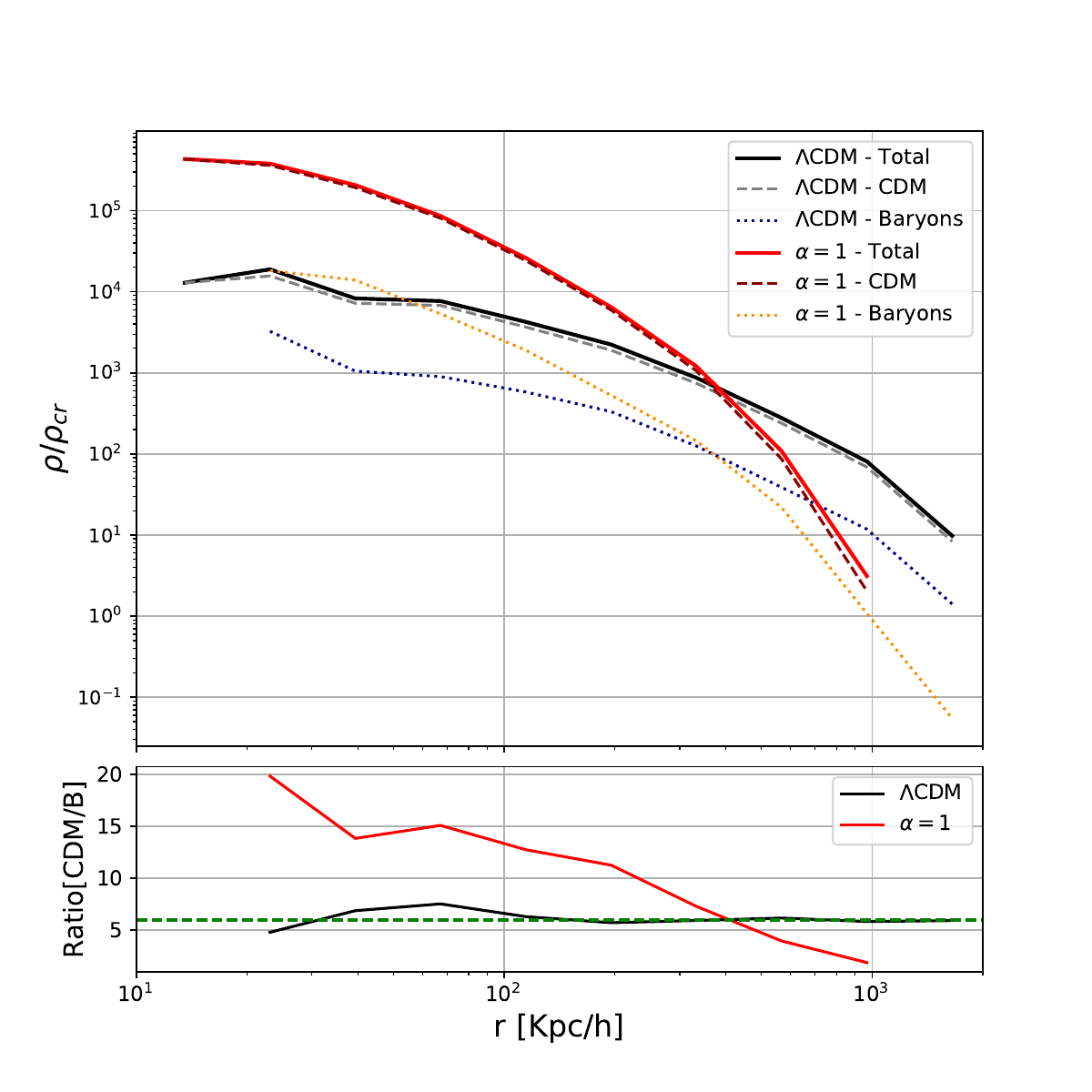}
\caption{{\it Top panels:} Stacked radial density profiles for a sample of 40 halos identified to be the same in both the $\Lambda$CDM simulation (black-blue lines) and the covariantised dark Thomson-like interacting model simulation (red-orange lines). We show the total (solid lines), dark matter (dashed lines) and baryonic (dotted lines) density profiles. \\
{\it Bottom panels:} ratio between the density profile of dark matter and baryons for the $\Lambda$CDM simulation (black lines) and the covariantised dark Thomson-like interacting model simulation (red lines) and the reference ratio $\frac{\Omega_{\rm dm}}{\Omega_{\rm b}}$ (green dashed line). We show the results for two different halo masses with reference to the non-interacting case, those whose halo mass is $M_{\rm halo} \sim 10^{15}\; M_\odot/h$  (left panel) and those whose halo mass is $M_{\rm halo} \sim 10^{14}\; M_\odot/h$ (right panel). Since we started from the same initial conditions, we identify each of the halos by being formed in the same spatial position with similar mass ranges, taking into account the covariantised dark Thomson-like scattering produces less massive halos.
}
\label{fig:profile}
\end{figure*}

From the previous section, we have discussed how the momentum transfer induced by the elastic interaction results in a noticeable suppression of very massive halos compared to a non-interacting scenario. However, a key question remains unanswered: to what extent does this interaction reshape the process of individual halo formation? This issue will be discussed next.

In Figure~\ref{fig:profile}, we present the averaged radial density profiles of a representative sample of the same halos found in both the $\Lambda$CDM scenario (black line) and the covariantised dark Thomson-like interacting model (red line). Let us recall that both simulations start from the same initial conditions, and that the clustering process evolves identically up to the onset of the strong elastic interaction epoch. Thus, there will be a population of analogous halos that will form at similar positions in both simulations and these are the ones we shall compare to study the effects of the interaction on the halo profile. The selected halos are divided into two different mass scales of the reference non-interacting case to facilitate the visualization: $M_{\rm halo} \sim 10^{15}\; M_\odot/h$  (left panel) and $M_{\rm halo} \sim 10^{14}\; M_\odot/h$ (right panel). These plots allow us to directly compare the internal structure of the halos under both scenarios. Furthermore, we have used the critical density of the Universe today $\rho_{\text{cr}}$ to normalise the radial density of each halo.

From the resulting density profiles, the first interesting observation is that halos tend to be more compact and exhibit smaller physical sizes in the interacting case as compared to their counterparts in the standard $\Lambda$CDM scenario. The cores of the halos formed in the interacting model feature significantly higher central densities, suggesting that the interaction leads to more concentrated mass distributions. This picture is further supported by the fact that the radial density distribution declines more steeply in the interacting model, resulting in less massive and more diffuse outer regions of the halos. These results are in agreement with the explanation given above to understand the enhancement of the power spectrum on non-linear scales, i.e., that the interaction reduces the kinetic energy of dark matter particles in virialised objects so they become more compact. We can also see in Figure~\ref{fig:profile} how the profile of baryons is slightly less compact because, although they do not interact directly with dark energy, they feel a reduced gravitational potential and that the effect is more prominent for more massive halos. Let us notice that a distinctive signature of the interacting model is precisely the different profiles of dark matter and baryons as shown in the lower panels. A consequence of having different profiles for baryons and dark matter is that the standard spherical collapse model where the mass within each shell is conserved will fail before than in the standard case. In other words, shell-crossing will occur before than expected from a pure $\Lambda$CDM. With this in mind, it may be interesting to revisit the halo-model based on spherical collapse for quasi-linear scales of \cite{Lague:2024sox} with a more complete spherical collapse model that includes baryons (as an additional collisionless and non-interacting component) and takes into account shell crossings from both components. This set up with two matter components one of which features an interaction that modifies the spherical collapse resembles the scenarios with charged dark matter studied in \cite{BeltranJimenez:2020tsl} so we may expect some qualitative similarities.

While in the non-interacting case both dark matter and baryon profiles scale equally with distance,  the interacting scenario shows an enhancement of the radial density of dark matter in the innermost shells of the halos in comparison with baryons, while the outskirts become lighter in terms of dark matter. This observation is consistent with the previous results. The interaction, which only couples to dark matter, inhibits the accretion of dark matter particles at late times. During the formation of the halos, this effect primarily impacts their outer regions. Consequently, when the interaction becomes efficient at late times, dark matter accretion nearly ceases, whereas baryons continue to fall into the weakened gravitational potential wells. It is noteworthy that this particular effect extends to halos of all scales, rather than being limited to the most massive ones, as observed in previous effects. Therefore, the momentum transfer interactions seem to sweep the outskirts of the halo of dark matter particles. 

A final conclusion that emerges from these results is the following: in momentum transfer scenarios of this kind, the use of baryon profiles as templates of the underlying dark matter field within individual halos carries an additional bias which has a radial dependence. However, the central question is whether baryons can still serve as reliable tracers of the overall density field or, in other words, whether galaxies will continue to reside at the bottom of dark matter potential wells. This will be confirmed in the next section through an analysis of the cosmic web.

\subsection{Cosmic web}
We now turn our attention to how the cosmic web is altered by the presence of the elastic interaction. Given that momentum transfer occurs exclusively in the dark matter component and not in baryons, a fundamental question arises: will both matter components remain interconnected? In other words, and in relation to actual observations, do galaxies still serve as reliable tracers of the dark matter velocity field, and do they continue to reside at the bottom of dark matter potential wells?\\

\begin{figure*}[t]
\centering
\includegraphics[width=17cm]{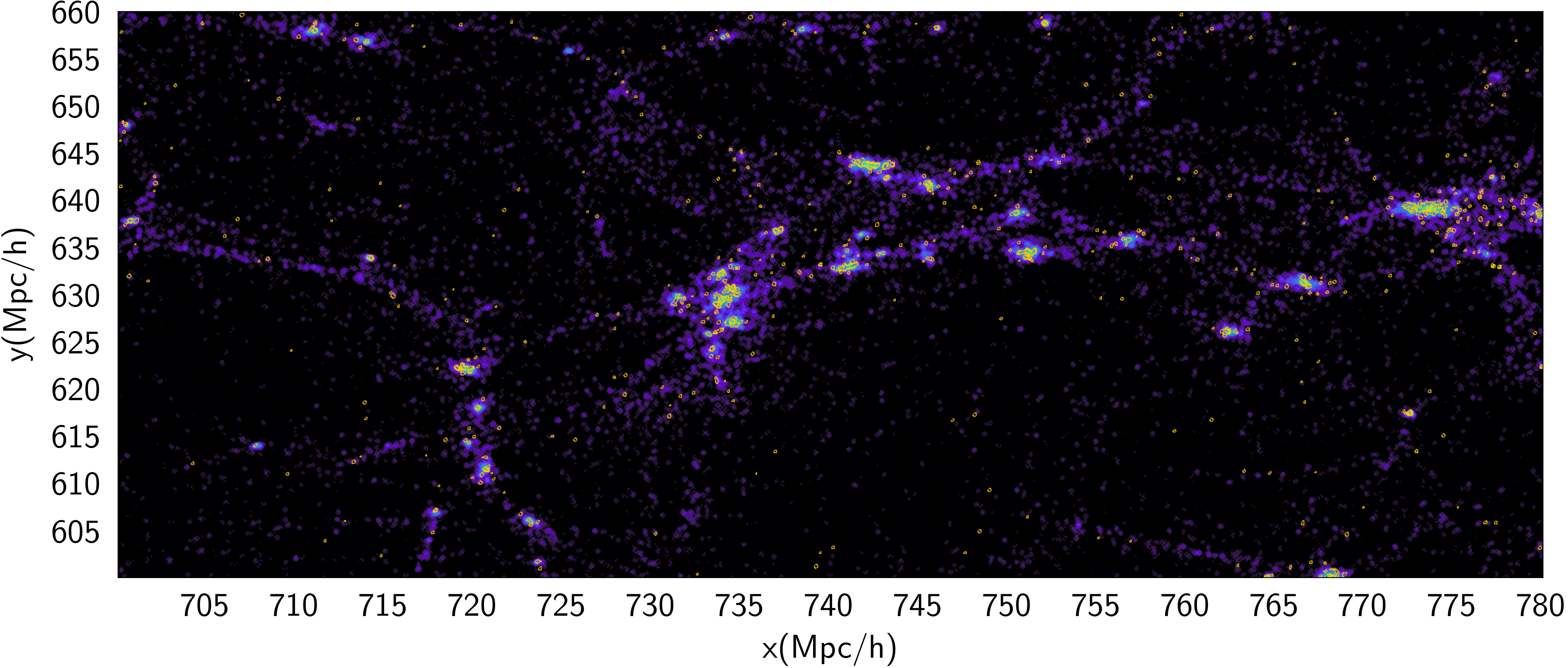}
\includegraphics[width=17cm]{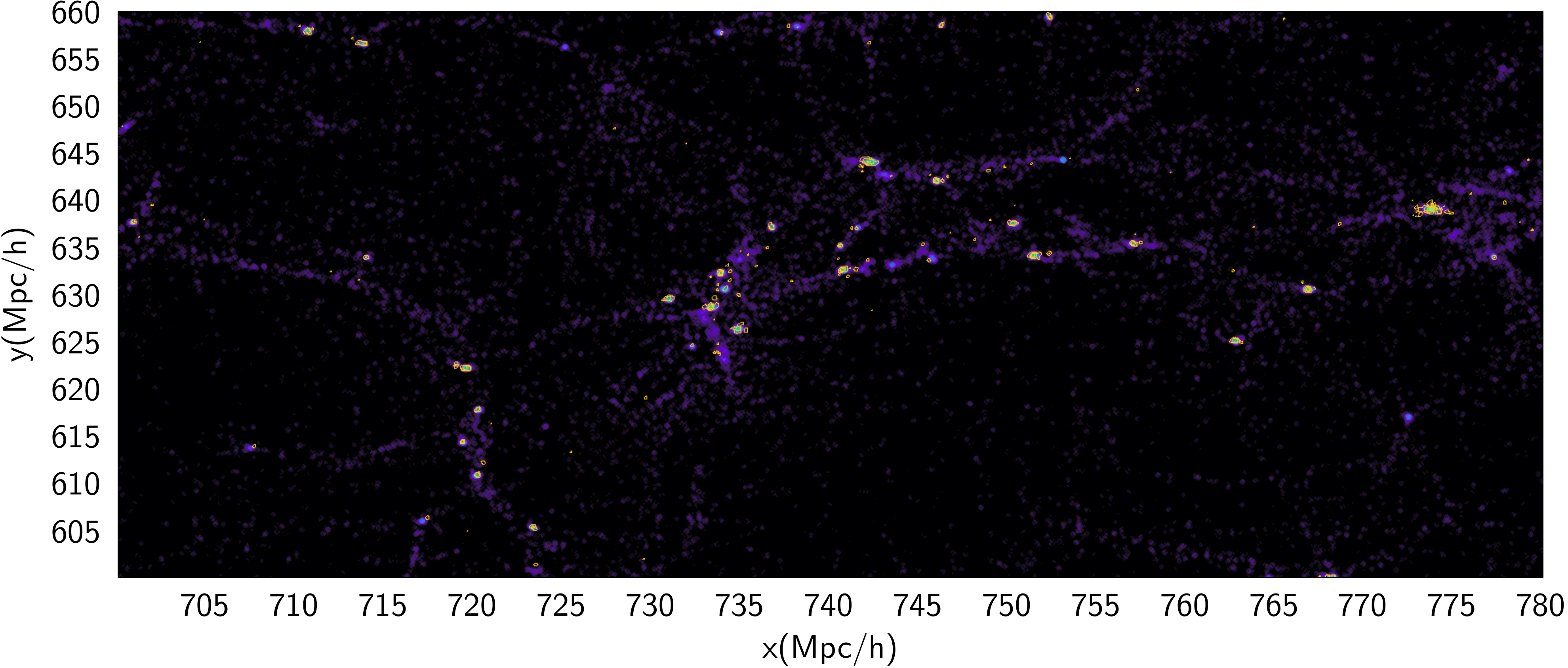}
\caption{Distribution of matter for $\Lambda$CDM (top) and the interacting model $\alpha$CDM (bottom). Dark matter is represented by the black-blue-green gradient density map while regions of high density of baryons are represented by yellow density lines. We confirm that the interaction gives rise to a less clustered universe and that baryons are still locked inside the dark matter halos.}
\label{fig:nbody_densitymap}
\end{figure*}

In Figure~\ref{fig:nbody_densitymap}, we show the dark matter distribution as a density field  ranging from black (representing lower dark matter density) to blue (indicating high concentrations of dark matter) for our choice of cosmological parameters and fiducial coupling parameter $\alpha=1$. 
Overlaid on this, we also display iso-density regions of high baryon concentration as yellow isogram lines, which mark the contours of constant baryon density. Contours corresponding to lower baryon densities are omitted for visualisation purposes. These contours effectively highlight the densest clumps of baryons in our simulations, corresponding to the locations where galaxies will reside.
A  weakening of the cosmic web is evident, as dark matter exhibits significantly reduced clustering in the interacting case when compared to the standard scenario. Furthermore, we note that regions of very high density,  both massive halo nodes and connecting filaments, appear more compact, reinforcing our previous analyses on how the elastic interaction clears of dark matter particles the surrounding areas of a halo, where in the standard scenarios accreting process is taking place.  

Regarding the previous questions on baryons still being tracers of dark matter, we infer from Figure~\ref{fig:nbody_densitymap} that baryons, that we take as proxies for galaxies in the simulations, remain gravitationally bound to dark matter halos. Galaxies are mainly formed prior to the beginning of the elastic interaction dominance. Once they have formed and virialised at the bottom of the potential wells shaped by dark matter halos, they remain attached to these structures, even though the interaction inhibits further halo growth. Baryon halos, our galaxies proxies, and their host halos constitute strongly bound systems, and for reasonable values of the coupling parameter, disassembly due to elastic interaction does not appear as seen in the Figure~\ref{fig:nbody_densitymap}. 
Taken together with the previous results, from the matter distribution we conclude that the overall distribution of baryons remains largely unaffected, which will still serve as a proxy of where galaxies will form. While from the profile analyses we infer the baryon halos will change due to the interaction, which suggest galaxy formation will suffer some kind of modification due to the interaction. However, we have to note that galaxies typically form and virialise before the onset of the elastic interaction and that its formation process is more complicated that our simulation where we just assigned some particles to be baryons. Thus, we conclude that once formed they will inhabit shallower dark matter potential wells due to the effect of the interaction on the surrounding dark matter distribution and, therefore, their accretion process will be reduced resulting in also, but less than halos, more compact objects.

\section{Discussion}
\label{sec:conclusion}
In this work, we have explored the non-linear regime of a particular example of the pure momentum transfer class of interactions in the dark sector, the so-called covariantised dark Thomson-like scattering between dark energy and dark matter. We have studied a fiducial model with the coupling parameter set to $\alpha = 1$, which corresponds to the best fit value obtained from cosmological probes in previous studies. The main goal here has been to explore small scales corresponding to the non-linear regime of structure formation. For this goal, we have implemented the model into the \texttt{RAMSES} code and performed two simulations, one for the standard $\Lambda$CDM scenario and the other for our interacting model. Furthermore, we have exploited that the interacting epoch commences at late times to set identical initial conditions for both simulations since the interaction is negligible at the initial redshift. This further means that noticeable effects only emerge at low redshift. The outputs of the simulations have shown that the elastic interaction gives rise to a suppression of the matter power spectrum for very late times and small scales when compared to $\Lambda$CDM. Thus, our results extend to the non-linear regime the suppression of clustering observed in previous works for linear scales and confirms the expected behaviour.

We have also analysed the formation of halos in the elastic interacting scenario and we have observed that very massive halos are significantly less likely to form than in the standard $\Lambda$CDM scenario, while less massive halos remain essentially unaffected in the amount of them that are formed. This feature has been confirmed both by studying the distribution of the amount of particles that belong to the halos and the distribution of the total mass of the halos found in the simulations.  This reflects the fact that most massive halos are the last to be formed in the hierarchical structure formation paradigm and, thus, they are the most affected by the interaction, which becomes efficient only at late times. We have to remember here that the only new parameter of the interaction, $\alpha$, is the one setting the time-scale when the interaction becomes efficient. Therefore, $\alpha$ also controls the threshold mass above which  the number of halos is reduced. For the case studied here, $\alpha=1$, it corresponds to masses around $2\times10^{14} M_{\odot}/h$ at $z=0.05$. Larger values of $\alpha$ will imply lighter halos affected, and vice versa for smaller values.
Thus, once the elastic interacting domination commences, the most massive structures at that moment, which would otherwise be merging and accreting during this period, will not. Less massive halos, on the other hand, were already formed prior to the interacting epoch in accordance to the hierarchical formation process of halo formation. Since the interaction is unable to disrupt those already formed halos, their amount is insensitive to the interaction.

On the other hand, when analysing the internal dynamics of individual halos, we have found that they undergo a noticeable increase in the steepness of the profiles. This implies that halos are more compact objects with shorter radial extension. This effect is particularly prominent in the coupled component, dark matter, which becomes increasingly concentrated in the inner regions of the halos, while the outskirts are emptied as the interaction suppresses further accretion. Driven by that effect on dark matter, baryons also suffer a similar but milder compactification of their profiles. Furthermore, inner layers of the halos show a significant enrichment of dark matter while outer layers contain a larger proportion of baryons as compared to $\Lambda$CDM. This occurs because the interaction prevents the accretion of dark matter into the potential wells, but does not act on baryons.  We have to note here that this effect, although more prominent for larger halos, affects all halos unlike the reduction of formed halos, which only affected the most massive halos to be formed when the interaction becomes efficient.
As a consequence, when tracing the underlying matter distribution with baryons an additional bias may have to be taken into account for halo scales.

Finally, we have also analysed how faithfully baryons trace the dark matter distribution. Although the interaction weakens the growth of structures in both filaments and nodes, we have found that high baryon density regions, that we assume to be a proxy for the location of galaxies, show a significant overlap with dark matter overdensities. This result suggests that baryons are still good tracers of the dark matter density field in the presence of the elastic interaction.

Recently, the interacting model considered in this work has been confronted to the latest ACT data in \cite{2025arXiv250314454C} with a modelling of the non-linear regime developed in \cite{Lague:2024sox}.\footnote{In Refs. \cite{2025arXiv250314454C} and \cite{Lague:2024sox}, it is stated that they consider the elastic scattering model of \cite{Simpson:2010vh}, although the considered model in those works is actually the covariantised model studied in this work. As explained above, although both scenarios formally give the same form of the linear perturbation equations, the crucial difference is that a proper Thomson scattering is more important at high redshift while the covariantised interacting model becomes more relevant at low redshift.} This non-linear modelling is based on a modified spherical collapse model for quasi-linear scales to obtain the critical overdensity for the elastic interacting model and an implementation into \texttt{HMCode}~\cite{Mead:2020vgs}. The $N$-body code developed in this paper will allow to test the validity of the method developed in \cite{Lague:2024sox} that was used in \cite{2025arXiv250314454C} and go beyond the quasi-linear scales. Incorporating other data sets in a fully consistent way requires a good modelling of the non-linear scales that were lacking in the literature. Our developed $N$-body code and our set of simulations fills this gap and will permit a more appropriate confrontation to data and assess to what extent current data prefer the elastic interaction. In this respect, Ref.~\cite{Doux:2025vru} has explored a scale-dependent modification of the power spectrum and shown that galaxy-lensing data from DESY3~\cite{DES:2022qpf}, KiDS-1000~\cite{KiDS:2020suj}, HSC Y3~\cite{Dalal:2023olq} and ACT DR6~\cite{ACT:2023kun,ACT:2023dou} prefer a 15-30\% suppression in the power spectrum as compared to Planck results. Since ACT lensing is consistent with Planck for $\Lambda$CDM and its sensitivity peaks at around $z\simeq2$, this further supports the idea that the effects may come from intermediate/small scales at low redshift (see discussion on this in e.g.~\cite{Jimenez:2024lmm}). The results of Ref.~\cite{Doux:2025vru} rely on the use of \texttt{HALOFIT} which, in principle may not be justified, but our results here show that it indeed gives a reasonably good account of the non-linear regime. On the other hand, it is interesting to note that the method developed in~\cite{Doux:2025vru} assumes a fixed background cosmology so that the elastic scenarios like the one considered here are specially well-suited since the background remains unaltered by construction. On the other hand, the scale-dependent modification of the power spectrum of the elastic model does depend on redshift and this would need to be included in the method of \cite{Doux:2025vru}. Given the simplicity of the elastic interaction effects, we expect the inclusion of these effects to be straightforward.

In conclusion, with these results we have tested the momentum transfer scenarios by checking the differences and assumptions in the formation of structures in the Universe and, in particular, regarding the upcoming weak lensing and galaxy-clustering data probes. These results also probe that despite the extremely simplicity of this scenario, with only one new parameter, several effects appear from halo formation process to global clustering and, therefore, forthcoming experiments will be able to set constraints in the momentum transfer models. An exhaustive exploitation of data from surveys like Euclid~\cite{2011arXiv1110.3193L}, DES~\cite{2005astro.ph.10346T}, DESI~\cite{DESI:2016fyo} or J-PAS~\cite{J-PAS:2014hgg} requires a comprehensive knowledge of the non-linear clustering to reliably extract information from the data\footnote{For instance, the recent forecast analysis performed in \cite{Cruickshank:2025iig} for the scenario studied in this work, but with a redshift-dependent interacting parameter, could be refined to e.g. include some quasi-linear or non-linear scales in a robust way.} and this will become even more important for the next generation of stage-5 experiments \citep[see e.g.][]{Spec-S5:2025uom}.  This work represents a first step in this direction. Future work will refine our results with higher resolution and complete simulations as well as analysing other promising observables to test the elastic interacting model. In this respect, it will be interesting to analyse velocity correlations since the interaction precisely depends on the peculiar velocities of dark matter. Another interesting observable might be looking at the properties of voids. It has been recently pointed out in \cite{Schuster:2025qco} that the evolution of voids in $\Lambda$CDM stabilises at redshift around $z\simeq1$, which is the redshift at which the elastic interaction becomes relevant according to the  obtained constraints in previous works~\cite{Figueruelo:2021elm}. Thus, we may expect the elastic interaction to generate further evolution of voids below redshift $z=1$ which could provide another distinctive signature of these scenarios.

\bibliographystyle{JHEP}
\bibliography{biblio}

\end{document}